%% file: main.tex
\documentclass[a4paper,reqno]{amsart}
\usepackage{amsfonts}
\usepackage{array,calc} 
\usepackage{amsmath,calc,graphicx}

\usepackage{url}
\usepackage[hidelinks]{hyperref}
\usepackage{amssymb}
\usepackage{amsthm}
\usepackage{float}
\usepackage{enumitem}
\usepackage{caption}
\usepackage{subcaption}
\usepackage{amsrefs}
\usepackage{todonotes}
\numberwithin{equation}{section}
\numberwithin{figure}{section}
\theoremstyle{plain}
\newtheorem{theorem}{Theorem}[section]
\newtheorem{lemma}[theorem]{Lemma}
\newtheorem{corollary}[theorem]{Corollary}
\newtheorem{proposition}[theorem]{Proposition}
\newtheorem{definition}[theorem]{Definition}

\theoremstyle{remark}
\newtheorem{remark}[theorem]{Remark}

\newtheorem*{lem*}{\textsc{Lemma}}
\newtheorem*{cor*}{\textsc{Corollary}}
\newtheorem*{exer*}{\textsc{Exercise}}
\newtheorem*{con*}{\textsc{Conjecture}}
\newtheorem*{thm*}{\textsc{Theorem}}
\newcommand{\beq}{\begin{equation}}
\newcommand{\eeq}{\end{equation}}

\newcommand\Pfour{\textrm{P}_{\textrm{IV}}}

\usepackage{chngcntr}
\counterwithout{figure}{section}

\newcommand{\orcidauthorA}{0000-0001-7504-4444}

\title{On symmetric solutions of the fourth $q$-Painlev\'e equation}
\date{\today}
\author{Nalini Joshi}
\thanks{NJ's ORCID ID is \orcidauthorA. NJ's research was supported by an
  Australian Research Council Discovery Project \#DP210100129.
}

\address{School of Mathematics and Statistics F07, University of Sydney, NSW 2006 Australia}
\email{nalini.joshi@sydney.edu.au}
\author{Pieter Roffelsen}
\email{pieter.roffelsen@sydney.edu.au}
  
\begin{document}
\begin{abstract}
The Painlev\'e equations possess transcendental solutions $y(t)$ with special initial values that are symmetric under rotation or reflection in the complex $t$-plane. They correspond to monodromy problems that are explicitly solvable in terms of classical special functions. In this paper, we show the existence of such solutions for a $q$-difference Painlev\'e equation. We focus on symmetric solutions of a $q$-difference equation known as $q\Pfour$ or $q{\rm P}(A_5^{(1)})$ and provide their symmetry properties and solve the corresponding monodromy problem.
\end{abstract}
\maketitle

\tableofcontents
\input{intro}

\input{symmetric_solutions}

\input{symmetric_monodromy}

\input{monodromy_solvable}

\input{meromorphic_sol}

\input{conclusion}
\appendix

\input{app_notation}
\input{app_technical_lem}

\end{document}

%% file: intro.tex
\section{Introduction}
Among the highly transcendental solutions $y(t)$ of a Painlev\'e equation, there exist solutions with solvable monodromy \cites{kitaev1991symmetrical,kaneko2005,pvisymmetric,okumura2007symmetric}, often called {\em symmetric solutions}. For generic parameter values, they are neither classical special functions\footnote{These are defined by Umemura as solutions related to hypergeometric-type or rational functions under classical transformations.}  \cite{umemura1998painleve} nor solutions characterized by distinctive asymptotic behaviours, such as the celebrated tritronqu\'ee solutions \cite{jk:01}. In this paper, we show that symmetric solutions also exist for $q$-difference Painlev\'e equations. 

To be explicit, we focus on the $q$-difference fourth Painlev\'e equation 
\begin{equation*}
\ q\Pfour(a):\begin{cases}\displaystyle
\frac{\overline{f}_0}{a_0a_1f_1}=\frac{1+a_2f_2(1+a_0f_0)}{1+a_0f_0(1+a_1f_1)}, &\\
\displaystyle\frac{\overline{f}_1}{a_1a_2f_2}=\frac{1+a_0f_0(1+a_1f_1)}{1+a_1f_1(1+a_2f_2)}, &\\
\displaystyle\frac{\overline{f}_2}{a_2a_0f_0}=\frac{1+a_1f_1(1+a_2f_2)}{1+a_2f_2(1+a_0f_0)}, &
\end{cases}
\end{equation*}
where $q\in\mathbb{C}$, $0<|q|<1$, is given, $f=(f_0,f_1,f_2)$ is a function of $t\in T\subseteq\mathbb{C}$ and $a:=(a_0,a_1,a_2)$ are constant parameters, subject to
\begin{equation}\label{eq:conditionfa}
f_0f_1f_2=t^2,\quad a_0a_1a_2=q,
\end{equation}
 $T$ is invariant under multiplication by $q$, and $\overline f=f(qt)$.
This equation is also known as $q{\rm P}(A_5^{(1)})$ in Sakai's diagram \cite{sakai2001}.

We will focus on  solutions of $q\Pfour(a)$ that are invariant under the following transformations. 
\begin{definition}\label{def:symmetry}
The following transformations are called {\em discrete symmetries} of $q\Pfour(a)$: 
\begin{equation}\label{eq:symmetry1}
\mathcal{T}_\pm:    t \mapsto \frac{\pm 1}{t},\quad (f_0,f_1,f_2)\mapsto (F_0,F_1,F_2)=(f_0^{-1},f_1^{-1},f_2^{-1}),
\end{equation}
i.e.,
\begin{equation*}
 F_k(t)=\frac{1}{f_k(\pm 1/t)}\quad (0\leq k\leq 2),
\end{equation*}
We call $T$ a {\em symmetric domain} if it is invariant under $t \mapsto \frac{\pm 1}{t}$. Furthermore, a solution $f$ of $q\Pfour(a)$ is called a {\em symmetric solution} if it is invariant under one of the above two symmetries.  \end{definition}
\noindent We show that $q\Pfour(a)$ is invariant under transformation 
\eqref{eq:symmetry1} in Section \ref{s:symm}. It is important to note that the above symmetries do not arise as elements of the affine Weyl symmetry group $(A_2+A_1)^{(1)}$ usually associated with $q\Pfour(a)$, but they turn out to correspond to one and the same automorphism of the corresponding Dynkin diagram. In particular, the symmetries are indistinguishable on the level of $q\Pfour(a)$, but they do act distinctively on the corresponding Lax pair, which we introduce next.

The difference equation $q\Pfour(a)$ is associated to a linear problem (called a {\em Lax pair}) \cite{joshinobu2016}
\begin{subequations}
	\label{eq:laxpair}
	\begin{align}\label{eq:laxspectral}
	Y(qz,t)&=A(z;t,f,u)Y(z,t),\\
	Y(z,qt)&=B(z;t,f,u)Y(z,t),\label{eq:laxtime}
	\end{align}
     \end{subequations}
     where $A$ and $B$ are matrix-valued functions given in Equations \eqref{eq:JNLax}. 
The compatibility condition
\begin{equation}\label{eq:ABcomp}
 A(z, qt)B(z, t)=B(qz,t)A(z,t),
  \end{equation}
is equivalent to the $q\Pfour(a)$ equation, along with a condition on the auxiliary variable $u$ given by
\begin{equation}\label{eq:auxiliaryuf}
	\frac{\overline{u}}{u}=b^2,
      \end{equation}
where $b$ is given by equation \eqref{eq:bb}. 

The linear problem \eqref{eq:laxspectral} gives rise to a Riemann-Hilbert problem (RHP). In a previous paper, we showed that this Riemann-Hilbert problem is uniquely solvable (under certain conditions) and proved the invertibility of the map between the linear problem  and an algebraic surface, which is a $q$-version of a monodromy surface \cite{joshiroffelsenrhp}. Necessary notation is outlined in Appendix \ref{app:not}.

% that remains invariant under the deformation given by \eqref{eq:laxtime}
%For completeness, we provide the detailed definitions underlying the RHP and required conditions in Appendix \ref{app:rhp}. 

The main result of this paper, Theorem \ref{thm:solvable}, shows that solutions that are symmetric with respect to $\mathcal{T}_-$ lead to an explicitly solvable monodromy problem at the point of reflection, with solutions built out of Jackson's $q$-Bessel functions of the second kind, $J_\nu(x;p)$, with $p=q^2$ and exponents $\nu=\pm \tfrac{1}{2}$.
The construction of the monodromy surface breaks down at reflection points for the case of $\mathcal{T}_+$, because it violates the non-resonance conditions of the Riemann-Hilbert problem.

For the special choice of the parameters, $a_0=a_1=a_2=q^{\frac{1}{3}}$, $q\Pfour$ has a particularly simple solution, given by
\begin{equation*}
    f_0=f_1=f_2=t^{\frac{2}{3}},
\end{equation*}
which is symmetric with respect to both $\mathcal{T}_+$ and $\mathcal{T}_-$. We show that the monodromy problem of this solution is solvable everywhere in the complex plane.
This solution forms a seed solution for the family of $q$-Okamoto rational solutions, introduced in Kajiwara et al. \cite{kajiwaranoumiyamada2001}. In this paper, we provide the points on the monodromy surface corresponding to each member of this family.

\subsection{Outline}
The symmetric solutions and their derivations are described in detail in Section \ref{s:symm}. The corresponding linear problem, connection matrix, and monodromy surface are considered in Section \ref{s:mono}. In Section \ref{sec:solve_linear}, we show that the monodromy problem for symmetric solutions is solvable at points of reflection. We consider symmetric solutions on open domains in Section \ref{s:mero}, particularly focussing on the $q$-Okamoto rational solutions, before providing a conclusion in Section \ref{s:conc}.

%% file: symmetric_solutions.tex
\section{Symmetric Solutions}\label{s:symm}
In this section, we first show that $q\Pfour$ remains invariant under the transformations given in Definition \ref{def:symmetry}.
Then, in Section \ref{subsec:continuum}, we show that the transformations formally converge to a transformation of the fourth Painlev\'e equation under the continuum limit. Finally, in Section \ref{sec:classify_symmetric}, we classify solutions, symmetric with respect to $\mathcal{T}_-$.

To show that $\mathcal{T}_{\pm}$ leave $q\Pfour$ invariant, note that these transformations map
\begin{equation}\label{eq:symmonf}
    f_k\mapsto 1/F_k,\quad \overline{f}_k \mapsto 1/\underline{F}_k,\quad  \underline{f}_k \mapsto 1/\overline{F}_k,\qquad (k=0,1,2).
\end{equation}
Taking $t\mapsto 1/t$ in $q\Pfour(a)$ we obtain
\begin{equation*}
\begin{cases}\displaystyle
\underline{f}_0=\displaystyle a_0^{-1}a_2^{-1}f_2\,\frac{1+a_1^{-1}f_1(1+a_0^{-1}f_0)}{1+a_0^{-1}f_0(1+a_2^{-1}f_2)}, &\\
\underline{f}_1=\displaystyle a_0^{-1}a_1^{-1}f_0\,\frac{1+a_2^{-1}f_2(1+a_1^{-1}f_1)}{1+a_1^{-1}f_1(1+a_0^{-1}f_0)}, &\\
\underline{f}_2=\displaystyle a_1^{-1}a_2^{-1}f_1\,\frac{1+a_0^{-1}f_0(1+a_2^{-1}f_2)}{1+a_2^{-1}f_2(1+a_1^{-1}f_1)}. &
\end{cases}
\end{equation*}
Using Equations \eqref{eq:symmonf} to replace lower-case variables by upper-case variables, we find another instance of $q\Pfour(a)$, with the same parameters.

Recall that $q\Pfour$ has a symmetry group given by $(A_2+A_1)^{(1)}$ (see \cite[\S 4]{joshinobushi2016}). We note here that the  transformations $\mathcal{T}_{\pm}$ are not given by the generators of the reflection group $(A_2+A_1)^{(1)}$, but are related to an automorphism of the corresponding Dynkin diagram. To be precise, they are equivalent to $r$ in \cite[\S 4.2]{joshinobushi2016}.

\subsection{$\mathcal{T}_{\pm}$ and the continuum limit}\label{subsec:continuum}
In Kajiwara et al. \cite{kajiwaranoumiyamada2001}, it was shown that, upon setting
\begin{align*}
    f_k(t,\epsilon)&=-\exp\left({-\epsilon g_k(s)+\mathcal{O}(\epsilon^2)}\right)\quad (k=0,1,2),\\
    t^2&=\exp\left(-\epsilon s\right),\\
    a_k&=\exp\left(-\tfrac{1}{2}\epsilon^2 \alpha_k\right)\quad (k=0,1,2),\\
    q&=\exp\left(-\tfrac{1}{2}\epsilon^2\right),
\end{align*}
and taking the limit $\epsilon\rightarrow 0$, $q\Pfour$ formally converges to the symmetric fourth Painlev\'e equation
\begin{equation*}
\ S\Pfour(\alpha):\begin{cases}
\displaystyle g_0'=\alpha_0+g_0(g_1-g_2), &\\
\displaystyle g_1'=\alpha_1+g_1(g_2-g_0), &\\
\displaystyle g_2'=\alpha_2+g_2(g_0-g_1), &\\
\end{cases}
\end{equation*}
where
\begin{equation*}
    g_0+g_1+g_2=s,\quad \alpha_0+\alpha_1+\alpha_2=1,
\end{equation*}
and $g'=g'(s)$ denotes differentiation with respect to $s$.

Note that the independent $t$ variable is given by
\begin{equation*}
    t=t(s;\epsilon)=\pm i\exp\left(-\epsilon s\right),
\end{equation*}
and satisfies
\begin{equation*}
    t(-s;\epsilon)=c/t(s;\epsilon),\quad c=\pm 1.
\end{equation*}

Thus, for $k=0,1,2$,
\begin{align*}
      F_k(t,\epsilon)&=1/f_k(c/t,\epsilon)\\
      &=-\exp\left({+\epsilon \,g_k(-s)+\mathcal{O}(\epsilon^2)}\right)\\
      &=-\exp\left({-\epsilon \,G_k(s)+\mathcal{O}(\epsilon^2)}\right),
\end{align*}
where
\begin{equation*}
    G_k(s)=-g_k(-s)\quad (k=0,1,2).
\end{equation*}

Therefore, in the continuum limit as $\epsilon\rightarrow 0$, the symmetries of $q\Pfour$ in Definition \ref{def:symmetry}, formally converge to the following symmetry of $S\Pfour$,
\begin{equation*}
s\rightarrow -s,\quad g_k\rightarrow G_k=-g_k\quad (k=0,1,2).
\end{equation*}

\subsection{Symmetric Solutions}\label{sec:classify_symmetric}
In this section, we restrict our attention to solutions with a domain given by a discrete $q$-spiral, $T=q^{\mathbb{Z}}t_0$.
For the symmetric transformations given in Definition \ref{def:symmetry}, we require that $t\rightarrow c/t$, $c=\pm 1$, leaves this spiral invariant.
This gives us four possible values for $t_0$, modulo $q^\mathbb{Z}$, determined by
\begin{equation*}
    t_0=c/t_0,\quad c=\pm 1,
\end{equation*}
namely $t_0=1,i,-1,-i$.

The formulation of the $q$-monodromy surface described in Section \ref{s:mono} requires the non-resonance conditions
\begin{equation}\label{eq:conditionnonresonant}
t_0^2,\pm a_0,\pm a_1,\pm a_2\notin q^\mathbb{Z}.
\end{equation}
This leads to two possible values, $t_0=\pm i$. As $q\Pfour(a)$ is invariant under $t\mapsto -t$, we restrict ourselves to considering $t_0=i$.

For any solution $f=f(q^m i)$, $m\in\mathbb{Z}$, of $q\Pfour(a)|_{t_0=i}$, the symmetry \eqref{eq:symmetry1} shows that
\begin{equation}\label{eq:symmetry_discrete}
    F_k(q^m i)=\frac{1}{f_k(q^{-m}i)},\quad (m\in\mathbb{Z},k=0,1,2),
\end{equation}
defines another solution of $q\Pfour(a)|_{t_0=i}$.
\begin{definition}
  We call a solution $f=f(q^m i)$, $m\in\mathbb{Z}$, of $q\Pfour(a)|_{t_0=i}$ {\em symmetric} if it is invariant under the transformation \eqref{eq:symmetry_discrete}, i.e. if
  \begin{equation}\label{eq:sym_condition}
    f_k(q^m i)=\frac{1}{f_k(q^{-m}i)},\quad (m\in\mathbb{Z},k=0,1,2).
\end{equation}
\end{definition}

Consider a symmetric solution $f=f(q^m i)$, $m\in\mathbb{Z}$. Specialising equation \eqref{eq:sym_condition} to $m=0$, shows that $v_k:=f_k(i)\in \mathbb{CP}^1$ satisfies
$v_k=1/v_k$, for $k=0,1,2$. The only solutions to this equation are given by $v_k=\pm 1$. Thus $f$ is regular at $t=i$ and
\begin{equation}\label{eq:initial_conditions_constraint}
    f_k(i)^2=1,\quad (k=0,1,2).
\end{equation}
Combining this observation with
\begin{equation*}
    f_0(i)f_1(i)f_2(i)=-1,
\end{equation*}
we are led to four possible initial conditions at $m=0$,
\begin{equation}\label{eq:symmic}
   (f_0(i),f_1(i),f_2(i))\in \{(-1,1,1),(1,-1,1),(1,1,-1),(-1,-1,-1)\}.
\end{equation}
Conversely, any of these initial conditions yields a symmetric solution of $q\Pfour(a)|_{t_0=i}$. To see this, recall that equation \eqref{eq:symmetry_discrete} yields, in general, another solution $F$ of $q\Pfour(a)|_{t_0=i}$. Due to \eqref{eq:initial_conditions_constraint}, $f$ and $F$ satisfy the same initial conditions at $m=0$. Therefore, they are the same solution and thus $f$ is a symmetric solution. This proves the following lemma.
\begin{lemma}\label{lem:symmetric_solutions}
$q\Pfour(a)|_{t_0=i}$ has precisely four symmetric solutions, which are all regular at $t=i$, each specified by its initial values at $m=0$, with the four possible initial conditions given by
\begin{equation*}
   (f_0(i),f_1(i),f_2(i))=\begin{cases} 
   (-1,1,1),\\
   (1,-1,1),\\
   (1,1,-1),\\
   (-1,-1,-1).
   \end{cases}
\end{equation*}
\end{lemma}

\begin{figure}[t]
\centering
\includegraphics[width=\textwidth]{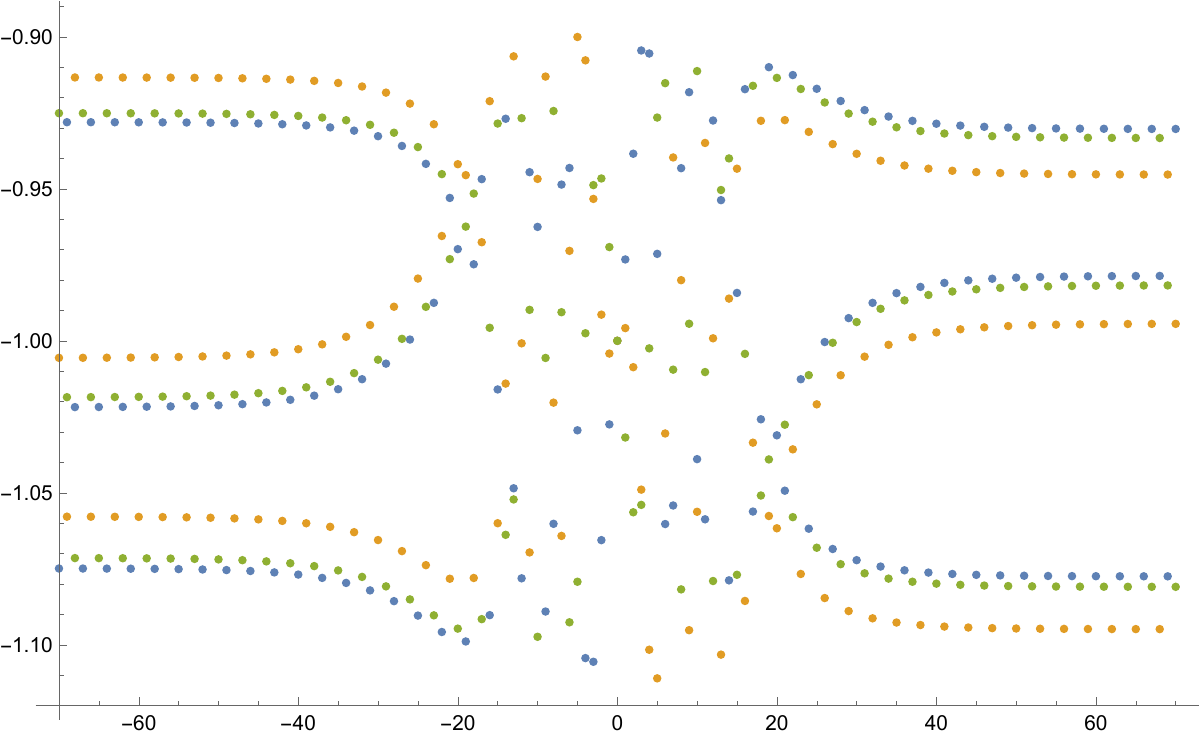}
    \caption{\label{fig:numerics}Numerical display of the symmetric solution in Lemma \ref{lem:symmetric_solutions} with initial conditions $(f_0(i),f_1(i),f_2(i))=(-1,-1,-1)$. The values of $q^{-\frac{2}{3} m}f_k(q^m i)$, $k=0,1,2$, are displayed in respectively blue, orange and green, with $m$ ranging from -70 to 70 on the horizontal axis. The values of the parameters are
    $a_0=q^{\frac{9}{23}}$, $a_1=q^{\frac{8}{23}}$ and $a_2=q^{\frac{6}{23}}$, with $q=0.802$.} 
\end{figure}

See Figure \ref{fig:numerics} for a plot of one the symmetric solutions.

\begin{remark}
It is instructive to compare this with the symmetric solutions of $S\Pfour(\alpha)$. In accordance with the definition of symmetric solutions of $\Pfour$, see Kaneko \cite{kaneko2005}, these are solutions $g$ of $S\Pfour(\alpha)$ that satisfy
\begin{equation*}
    g_k(s)=-g_k(-s)\quad (k=0,1,2).
\end{equation*}
$S\Pfour(\alpha)$ has precisely four symmetric solutions. 
Three non-analytic at $s=0$, with Laurent series in a domain around $s=0$ given by
\[ {\rm Case\ I:}\quad 
\begin{cases}
g_0(s)&=-\alpha_0 s+\mathcal{O}\left(s^3\right),\\
g_1(s)&=+s^{-1}+\mathcal{O}\left(s\right),\\
g_2(s)&=-s^{-1}+\mathcal{O}\left(s\right),
\end{cases}
\]
\[ {\rm Case\ II:}\quad 
\begin{cases}
g_0(s)&=-s^{-1}+\mathcal{O}\left(s\right),\\
g_1(s)&=-\alpha_1 s+\mathcal{O}\left(s^3\right),\\
g_2(s)&=+s^{-1}+\mathcal{O}\left(s\right),
\end{cases}
\]
\[ {\rm Case\ III:}\quad 
\begin{cases}
g_0(s)&=+s^{-1}+\mathcal{O}\left(s\right),\\
g_1(s)&=-s^{-1}+\mathcal{O}\left(s\right),\\
g_2(s)&=-\alpha_2 s+\mathcal{O}\left(s^3\right),
\end{cases}
\]
and one analytic at $s=0$, specified by
\begin{equation*}
{\rm Case\ IV:}\quad 
    g_k(s)=\alpha_k s+\mathcal{O}\left(s^3\right)\quad (s\rightarrow 0),
\end{equation*}
for $k=0,1,2$.
\end{remark}

%% file: symmetric_monodromy.tex
\section{Symmetries and the linear problem}\label{s:mono}
In this section, we recall some essential aspects of the linear problem associated with $q\Pfour$ and study their interplay with the symmetries $\mathcal{T}_{\pm}$.

In Section \ref{subsection:recaplinearproblem} we recall the Lax pair associated with $q\Pfour$ and lift the action of $\mathcal{T}_{\pm}$ to it. Then, in Section \ref{eq:symmetric_connect}, we introduce the connection matrix associated with the linear problem and derive how the symmetries act on it. Finally, in Section \ref{subsec:tyurin}, we compute how $\mathcal{T}_{\pm}$ transform certain monodromy coordinates and provide an alternative way to classify symmetric solutions.

\subsection{The Lax pair}\label{subsection:recaplinearproblem}
We recall the following Lax pair of $q\Pfour$, derived in \cite{joshinobu2016},
\begin{subequations}\label{eq:qLax}
  \begin{align}
  Y(q z, t)&=A(z, t)Y(z,t),\label{eq:A}\\
  Y(z, q t)&=B(z, t)Y(z,t),\label{eq:B}
  \end{align}
\end{subequations}
where 
\begin{subequations}\label{eq:JNLax}
\begin{align}
\notag A:=&\begin{pmatrix}
u & 0\\
0 & 1
\end{pmatrix}\begin{pmatrix}
-i\,q\frac{t}{f_2}z & 1\\
-1 & -\,i\,q\frac{f_2}{t} z
\end{pmatrix}
\begin{pmatrix}
-\,i\,a_0a_2\frac{t }{f_0}z & 1\\
-1 & -\,i\,a_0a_2\frac{f_0}{t} z
\end{pmatrix}\times\\
&\ \times\,
\begin{pmatrix}
-\,i\,a_0\frac{t}{f_1}z & 1\\
-1 & -\,i\,a_0\frac{f_1}{t} z
\end{pmatrix}\begin{pmatrix}
u^{-1} & 0\\
0 & 1
\end{pmatrix},\label{eq:AJN}\\
B:=&\begin{pmatrix}
0 & -bu\\
b^{-1}u^{-1} & 0
\end{pmatrix}+
 \begin{pmatrix}
z & 0\\
0 & 0
\end{pmatrix}\,,\label{eq:BJN}
\end{align}
\end{subequations}
with
\begin{equation}\label{eq:bb}
b=\frac{t(1+a_1f_1(1+a_2f_2))}{i\,(qt^2-1)f_2}.
\end{equation}
We refer to the first equation of the Lax pair, equation \eqref{eq:A}, as the spectral equation. 

Compatibility of the Lax pair,
\begin{equation}\label{eq:comp}
    A(z,qt)B(z,t)=B(qz,t)A(z,t),
\end{equation}
is equivalent to $(f_0,f_1,f_2)$ satisfying $q\Pfour(a)$ and $u$ satisfying the auxiliary equation
\begin{equation}\label{eq:auxiliary}
    \frac{\overline{u}}{u}=b^2.
\end{equation}

We proceed to lift the symmetries $\mathcal{T}_{\pm}$ to this Lax pair. To this end, the following notation will be helpful. For any $2\times 2$ matrix $U$, we let $U^{\diamond}$ denotes the co-factor matrix, or adjugate transpose, of $U$.
In other words,
\begin{equation*}
    \begin{pmatrix}
    a & b\\
    c & d
    \end{pmatrix}^{\diamond}=\begin{pmatrix}
    d & -c\\
    -b & a
    \end{pmatrix}.
\end{equation*}
We further remind the reader that some of the notation used in this paper,  is outlined in Appendix \ref{app:not}.
\begin{lemma}
The symmetry $\mathcal{T}_+$ extends to the following symmetry of the Lax pair,
\begin{align*}
    Y(z,t)&\mapsto \widetilde{Y}(z,t)=Y^\diamond(z,1/t),\\
    A(z,t)&\mapsto \widetilde{A}(z,t)=A^\diamond(z,1/t),\\
    B(z,t)&\mapsto \widetilde{B}(z,t)=B^T(z,1/(qt)),
\end{align*}
and, consequently,
\begin{equation*}
    u(t)\mapsto \widetilde{u}(t)=\frac{1}{u(1/t)},\quad b(t)\mapsto \widetilde{b}(t)=-b(1/(qt)).
\end{equation*}
Similarly, the symmetry $\mathcal{T}_-$ extends to the following symmetry of the Lax pair,
\begin{align*}
Y(z,t)&\mapsto \widetilde{Y}(z,t)=r(z)\sigma_3Y^\diamond(z,-1/t),\\
    A(z,t)&\mapsto \widetilde{A}(z,t)=-\sigma_3 A^\diamond(z,-1/t)\sigma_3,\\
    B(z,t)&\mapsto \widetilde{B}(z,t)=\sigma_3B^T(z,-1/(qt))\sigma_3,
\end{align*}
where $r(z)$ any function that satisfies $r(qz)=-r(z)$,
and, consequently,
\begin{equation*}
    u\mapsto \widetilde{u}(t)=\frac{1}{u(-1/t)},\quad b(t)\mapsto \widetilde{b}(t)=b(-1/(qt)).
\end{equation*}
\end{lemma}
\begin{proof}
We only prove the extension of the first symmetry. The other one follows analogously.

Let us denote $A(z,t)=\mathcal{A}(z,t,f_0,f_1,f_2,u)$ and $B(z,t)=\mathcal{B}(z,t,b,u)$ and
consider the transformation
\begin{equation*}
    \mathcal{T}:Y(z,t)\mapsto \widetilde{Y}(z,t)=Y^\diamond(z,1/t).
\end{equation*}
This transformation induces the following action on the Lax matrices,
\begin{align*}
    A(z,t)&\mapsto \widetilde{A}(z,t)=A^\diamond(z,1/t),\\
    B(z,t)&\mapsto \widetilde{B}(z,t)=B^T(z,1/(qt)).
\end{align*}
As $(UV)^\diamond=U^\diamond V^\diamond$, it follows that
\begin{align*}
    A^\diamond(z,1/t)&=\mathcal{A}^\diamond(z,1/t,f_0(1/t),f_1(1/t),f_2(1/t),u(1/t))\\
    &=\mathcal{A}(z,t,F_0(t),F_1(t),F_2(t),\widetilde{u}(t)),
\end{align*}
with
\begin{equation*}
   \widetilde{u}(t)=\frac{1}{u(1/t)},\quad  F_k(t)=\frac{1}{f_k(1/t)}\quad (k=0,1,2).
\end{equation*}
Note that this is consistent with the symmetry $\mathcal{T}_+$, so that $\mathcal{T}$ indeed defines an extension of $\mathcal{T}_+$.

It remains to be checked that the action of $\mathcal{T}$ of $B(z,t)$ is consistent with its action on $A(z,t)$. That is, we need to ensure that
\begin{equation}\label{eq:bconsistent}
    \mathcal{B}^T(z,1/(qt),b(1/(qt)),u(1/(qt)))=\mathcal{B}(z,t,\widetilde{b}(t),\widetilde{u}(t)),
\end{equation}
where, in acccordance with equation \eqref{eq:bb},
\begin{equation*}
\widetilde{b}(t)=\frac{t(1+a_1F_1(t)(1+a_2F_2(t))}{i\,(qt^2-1)F_2(t)}.
\end{equation*}

Now, equation \eqref{eq:bconsistent} holds if and only if
\begin{equation*}
    \widetilde{b}(t)\widetilde{u}(t)=-\frac{1}{b(1/(qt)),u(1/(qt))}.
\end{equation*}
By substituting the expression for $\widetilde{u}(t)$, it follows that this is equivalent to
\begin{equation*}
    \widetilde{b}(t)=-\frac{u(1/t)}{b(1/(qt)),u(1/(qt))}.
\end{equation*}
By the auxiliary equation \eqref{eq:auxiliary}, we have $b^2=\overline{u}/u$, which simplifies the right-hand side, so that the identify to prove simply reads
\begin{equation*}
    \widetilde{b}(t)=-b(1/(qt)).
\end{equation*}
The last equality follows by direct computation, using the $q\Pfour$ time-evolution equations.

Finally, we note that the transformation $\mathcal{T}$ preserves the compatibility condition of the Lax pair \eqref{eq:comp}, which reaffirms the fact that $(F_0,F_1,F_2)$ is another solution of $q\Pfour$, and further shows that $\widetilde{u}$ solves the corresponding auxiliary equation.
\end{proof}

Now, consider any symmetric solution of $q\Pfour$ with respect to $\mathcal{T}_-$, then we can choose a corresponding solution $u$ of the auxiliary equation such that the Lax matrices have the symmetries
\begin{align*}
  A(z,t)&=-\sigma_3 A^\diamond(z,-1/t)\sigma_3,\\
    B(z,t)&=\sigma_3B^T(z,-1/(qt))\sigma_3.
\end{align*}
By specialising the first equation to $t=i$, we then find
\begin{equation}\label{eq:symmetric_linear}
      A(z,i)=-\sigma_3 A^\diamond(z,i)\sigma_3.
\end{equation}
This provides another way to classify the symmetric solutions of $q\Pfour(a)|_{t_0=i}$, by computing all the coefficient matrices $A(z,i)$ that possess the symmetry \eqref{eq:symmetric_linear}.

\subsection{The connection matrix}
In this section, we introduce the connection matrix associated with the Lax pair and deduce how the symmetries $\mathcal{T}_{\pm}$ act on it.

Firstly, we introduce a canonical solution at $z=\infty$ in the following lemma.
\begin{lemma}\label{lem:solinf}[Lemma 3.3 in \cite{joshiroffelsenrhp}] For any fixed $t$, there exists a unique $2\times 2$ matrix  $\Phi_\infty(z,t)$, meromorphic in $z$ on $\mathbb{C}^*$, such that
	\begin{align}\label{eq:Phiinfeqiv}
	\Phi_\infty(qz,t)&=\frac{1}{qa_0^2a_2i} z^{-3}A(z,t)\Phi_\infty(z,t)\begin{pmatrix}
	t^{-1} & 0\\
	0 & t
	\end{pmatrix},\\
	\Phi_\infty(z,t)&=I+\mathcal{O}\left(z^{-1}\right)\quad (z\rightarrow \infty).\label{eq:Phiinfeqiv2}
	\end{align}
In particular,
\begin{equation*}
Y_\infty(z,t)=\Phi_\infty(z,t)\begin{pmatrix}
r_+(z,t) & 0\\
0 & r_-(z,t)
\end{pmatrix}
\end{equation*}
defines a solution of the spectral equation \eqref{eq:A}, for any choice of functions $r_{\pm}(z,t)$ satisfying
\begin{equation*}
\frac{r_{\pm}(qz,t)}{r_{\pm}(z,t)}=qa_0^2a_2iz^{-3} t^{\pm 1}.
\end{equation*}
\end{lemma}

\begin{lemma}\label{lem:solzero}[Lemma 3.2 in \cite{joshiroffelsenrhp}] For any fixed $t$ and $d\in\mathbb{C}^*$, we have
	\begin{equation}\label{eq:A0iv}
	A(0)=M_0 \begin{pmatrix}
	i & 0\\
	0 & -i
	\end{pmatrix}M_0^{-1}, \textrm{ where } M_0:=d\begin{pmatrix}
	u & 0\\
	0 & 1
	\end{pmatrix}\cdot\begin{pmatrix}
	i & -i\\
	1 & 1
	\end{pmatrix},
	\end{equation}	
and, there exists a unique $2\times 2$ matrix $\Phi_0(z,t)$, meromorphic in $z$ on $\mathbb{C}^*$, such that
	\begin{align}\label{eq:Phizeroeqiv}
	\Phi_0(qz,t)&=A(z,t)\Phi_0(z,t)\begin{pmatrix}
	-i & 0\\
	0 & i
	\end{pmatrix},\\
	\Phi_0(z,t)&=M_0+\mathcal{O}\left(z\right),\  {\rm as}\ z\rightarrow 0.\nonumber
	\end{align}
In particular, it follows that
	\begin{equation*}
	Y_0(z,t)=\Phi_0(z,t)r_0(z)^{\sigma_3},
	\end{equation*}
	defines a solution of the spectral equation \eqref{eq:A}, for any choice of meromorphic function $r_0(z)$ satisfying $r_0(qz)=i\,r_0(z)$.
\end{lemma}

We define the corresponding connection matrix by
\begin{equation}\label{eq:assodefiiv}
C(z,t)=\Phi_0(z,t)^{-1}\Phi_\infty(z,t),
\end{equation}
which satisfies, see \cite{joshiroffelsenrhp}, for fixed $t$,
  \begin{enumerate}[label={{\rm (c.\arabic*)}},ref=c.\arabic*]
		\item $C(z,t)$ is analytic in $z$ on $\mathbb{C}^*$;\label{item:c1}
		\item $C(qz,t)=\frac{1}{qa_0^2a_2}z^{-3}\sigma_3C(z,t)t^{-\sigma_3}$;\label{item:c2}
		\item $|C(z,t)|=c\,\theta_q(a_0z,-a_0z,a_0a_2z,-a_0a_2z,qz,-qz)$, for some $c\neq 0$;\label{item:c3}
		\item $C(-z,t)=-\sigma_1C(z,t)\sigma_3$.\label{item:c4}
    \end{enumerate}

It follows from the compatibility condition \eqref{eq:comp}, see \cite{joshiroffelsenrhp} for more details, that
\begin{align*}
    \Phi_\infty(z,qt)&=B(z,t)\Phi_\infty(z,t)z^{-\sigma_3},\\
    \Phi_0(z,qt)&=B(z,t)\Phi_0(z,t)\sigma_3,
\end{align*}
which yields the almost trivial time-evolution of the connection matrix,
\begin{equation}\label{eq:connection_evolution}
    C(z,qt)=\sigma_3C(z,t)z^{-\sigma_3},
\end{equation}
as well as the time-evolution of $d$ in Lemma \ref{lem:solzero},
\begin{equation}\label{eq:timeevolutiond}
    \frac{\overline{d}}{d}=\frac{i}{b}.
\end{equation}

The connection matrix encompasses the monodromy of the Lax pair. In particular, one can in principle uniquely reconstruct the linear system \eqref{eq:A} from the connection matrix by solving an associated Riemann-Hilbert problem.

We will now extend the action of the symmetries to the connection matrix.
\begin{lemma}
The transformation $\mathcal{T}_+$ extends to the following symmetry of the canonical solutions and connection matrix,
\begin{align*}
    \Phi_\infty(z,t)&\mapsto \widetilde{\Phi}_\infty(z,t)=\Phi_\infty^{\diamond}(z,1/t),\\
    \Phi_0(z,t)&\mapsto \widetilde{\Phi}_0(z,t)=-i\,\Phi_0^{\diamond}(z,1/t)\sigma_1,\\
    C(z,t)&\mapsto \widetilde{C}(z,t)=i\,\sigma_1C^{\diamond}(z,1/t).
\end{align*}
The transformation $\mathcal{T}_-$ extends to the following symmetry of the canonical solutions and connection matrix,
\begin{align*}
    \Phi_\infty(z,t)&\mapsto \widetilde{\Phi}_\infty(z,t)=\sigma_3\Phi_\infty^{\diamond}(z,-1/t)\sigma_3,\\
    \Phi_0(z,t)&\mapsto \widetilde{\Phi}_0(z,t)=i\,\sigma_3\Phi_0^{\diamond}(z,-1/t),\\
    C(z,t)&\mapsto \widetilde{C}(z,t)=-i\,C^{\diamond}(z,1/t)\sigma_3.
\end{align*}
Furthermore, $\mathcal{T}_{\pm}$ act on $d$, defined in Lemma \ref{lem:solzero}, by
\begin{equation*}
    d(t)\mapsto \widetilde{d}(t)=d(\pm 1/t)u(\pm 1/t).
\end{equation*}
\end{lemma}
\begin{proof}
We only prove the extension for $\mathcal{T}_-$. The extension of $\mathcal{T}_+$ is proven analogously.

We first consider the canonical solution at $z=\infty$. In fact, by Lemma \ref{lem:solinf}, the matrix function $\Phi_\infty(z,t)$ is defined uniquely as the solution to \eqref{eq:Phiinfeqiv} and \eqref{eq:Phiinfeqiv2}. This means that the action of $\mathcal{T}_-$ on $\Phi_\infty(z,t)$ is already fixed by its action on the Lax matrix $A(z,t)$. 

To determine it explicitly, we first apply $t\mapsto -1/t$ to equation \eqref{eq:Phiinfeqiv}, which yields
\begin{equation*}
    \Phi_\infty(qz,-1/t)=-\frac{1}{qa_0^2a_2i} z^{-3}A(z,-1/t)\Phi_\infty(z,-1/t)\begin{pmatrix}
	t & 0\\
	0 & t^{-1}
	\end{pmatrix}.
\end{equation*}
Next, applying $U\mapsto U^{\diamond}$ to both sides, we obtain
\begin{equation*}
    \Phi_\infty^{\diamond}(qz,-1/t)=-\frac{1}{qa_0^2a_2i} z^{-3}A^{\diamond}(z,-1/t)\Phi_\infty^{\diamond}(z,t)\begin{pmatrix}
	t^{-1} & 0\\
	0 & t
	\end{pmatrix}.
\end{equation*}
Finally, multiplying both sides from the left and right by $\sigma_3$, we obtain
\begin{equation*}
    \widetilde{\Phi}_\infty(qz,t)=\frac{1}{qa_0^2a_2i} z^{-3}\widetilde{A}(z,t)\widetilde{\Phi}_\infty(z,t)\begin{pmatrix}
	t^{-1} & 0\\
	0 & t
	\end{pmatrix},
\end{equation*}
with
\begin{equation*}
    \widetilde{A}(z,t)=-\sigma_3 A^\diamond(z,-1/t)\sigma_3,\qquad \widetilde{\Phi}_\infty(z,t)=\sigma_3\Phi_\infty^{\diamond}(z,-1/t)\sigma_3.
\end{equation*}
Note that, furthermore, the normalisation at $z=\infty$ is correct, namely $\widetilde{\Phi}_\infty(z,t)=I+\mathcal{O}(z^{-1})$ as $z\rightarrow \infty$. We conclude, from Lemma \ref{lem:solinf}, that $\mathcal{T}_-$ indeed sends $\Phi_\infty(z,t)$ to $\widetilde{\Phi}_\infty(z,t)$.

We next consider the canonical solution at $z=0$. The matrix function $\Phi_0(z)$, see Lemma \ref{lem:solzero}, is only rigidly defined up to the choice of a scalar $d=d(t)$ which satisfies
$\overline{d}/d=i/b$, see equation \eqref{eq:timeevolutiond}. So, in order to fix the action of the symmetry $\mathcal{T}_-$ on $\Phi_0(z)$, we first need to fix its action on $d$ in such a way that $\overline{d}/d=i/b$ remains to hold true. Namely, it is required that, if we let $d\mapsto \widetilde{d}$ under $\mathcal{T}_-$, then
\begin{equation*}
    \frac{\widetilde{d}(qt)}{\widetilde{d}(t)}=\frac{i}{\widetilde{b}(t)}=-\frac{i}{b(-1/(qt))}.
\end{equation*}
We therefore set $\widetilde{d}(t)=d(-1/t)u(-1/t)$, so that indeed
\begin{equation*}
    \frac{\widetilde{d}(qt)}{\widetilde{d}(t)}=\frac{d(-1/(qt))}{d(-1/t)}\frac{u(-1/(qt))}{u(-1/t)}=\frac{b(-1/(qt))}{i}\frac{1}{b(-1/(qt))^2}=-\frac{i}{b(-1/(qt))}.
\end{equation*}

By essentially repeating the computation for $\Phi_\infty(z)$ above, for $\Phi_0(z)$, one finds that
\begin{equation*}
    \widetilde{\Phi}_0(z,t)=i\,\sigma_3\Phi_0^{\diamond}(z,-1/t),
\end{equation*}
defines a solution to, see equation \eqref{eq:Phizeroeqiv},
\begin{equation*}
    \widetilde{\Phi}_0(qz,t)=\widetilde{A}(z,t)\widetilde{\Phi}_0(z,t)\begin{pmatrix}
	-i & 0\\
	0 & i
	\end{pmatrix}.
\end{equation*}
Furthermore, direct evaluation of $ \widetilde{\Phi}_0(z,t)$ at $z=0$ gives
\begin{align*}
    \widetilde{\Phi}_0(0,t)&=i\,\sigma_3\Phi_0^{\diamond}(0,-1/t),\\
    &=i\, d(-1/t) \sigma_3 \begin{pmatrix}
	1 & 0\\
	0 & u(-1/t)
	\end{pmatrix}\cdot\begin{pmatrix}
	1 & -1\\
	i & i
	\end{pmatrix},\\
	&=d(-1/t) \begin{pmatrix}
	1 & 0\\
	0 & u(-1/t)
	\end{pmatrix}\cdot\begin{pmatrix}
	i & -i\\
	1 & 1
	\end{pmatrix}\\
	&=\widetilde{d}(t)\begin{pmatrix}
	\widetilde{u}(t) & 0\\
	0 & 1
	\end{pmatrix}\cdot\begin{pmatrix}
	i & -i\\
	1 & 1
	\end{pmatrix}.
\end{align*}
It follows that $\mathcal{T}_-$ sends $\Phi_0(z,t)$ to $\widetilde{\Phi}_0(z,t)$.

Finally, we compute the action of $\mathcal{T}_-$ on the connection matrix. Since $U\mapsto U^{\diamond}$ commutes with inversion, $U\mapsto U^{-1}$, we have
\begin{align*}
    \widetilde{C}(z,t)&=\widetilde{\Phi}_0(z,t)^{-1}\widetilde{\Phi}_\infty(z,t)\\
    &=\left[i\,\sigma_3\Phi_0^{\diamond}(z,-1/t)\right]^{-1}\sigma_3\Phi_\infty^{\diamond}(z,-1/t)\sigma_3\\
    &=-i\left[\Phi_0^{\diamond}(z,-1/t)\right]^{-1}\Phi_\infty^{\diamond}(z,-1/t)\sigma_3\\
    &=-i\, C^{\diamond}(z,-1/t)\sigma_3.
\end{align*}
This finishes the proof of the lemma.
\end{proof}

Now, let us take any symmetric solution of $q\Pfour$ with respect to $\mathcal{T}_-$, then we can choose a corresponding solution $u$ of the auxiliary equation, as well as $d$ satisfying \eqref{eq:timeevolutiond}, such that the connection matrix has the symmetry
\begin{equation*}
C(z,t)=-i\, C^{\diamond}(z,-1/t)\sigma_3.
\end{equation*}
By specialising this equation to $t=i$, we then find
\begin{equation}\label{eq:symmetric_connect}
      C(z,i)=-i\, C^{\diamond}(z,i)\sigma_3.
\end{equation}
This provides yet a third way to classify symmetric solutions of $q\Pfour(a)|_{t_0=i}$, by classifying all connection matrices $C(z,i)$ with the symmetry \eqref{eq:symmetric_connect}.

\subsection{Monodromy coordinates}\label{subsec:tyurin}
In \cite{joshiroffelsenrhp}, we introduced a set of coordinates on the connection matrix, which are invariant under right-multiplication of the connection matrix by diagonal matrices. They are given by
\begin{equation*}
\rho_k(t)=\pi(C(x_k,t)),\quad (1\leq k\leq 3),\quad (x_1,x_2,x_3)=(a_0^{-1},a_1/q,q^{-1}),
\end{equation*}
where, for any rank one $2\times2$ matrix $R$, letting $r_1$ and $r_2$ be respectively its first and second row,  $\pi(R)\in\mathbb{CP}^1$ is defined by
\begin{equation*}
r_1=\pi(R)r_2.
\end{equation*}
This yields three coordinates, $\rho=(\rho_1,\rho_2,\rho_3)\in(\mathbb{CP}^1)^3$, which satisfy the cubic equation,
\begin{align}\label{eq:cubic}
0=&+\beta_0\left[\theta_q(t)\rho_1\rho_2\rho_3-\theta_q(-t)\right]\\
&-\beta_1\left[\theta_q(t)\rho_1-\theta_q(-t)\rho_2\rho_3\right]\nonumber\\
&+\beta_2\left[\theta_q(t)\rho_2-\theta_q(-t)\rho_1\rho_3\right]\nonumber\\
&-\beta_3\left[\theta_q(t)\rho_3-\theta_q(-t)\rho_1\rho_2\right].\nonumber
\end{align}
with coefficients given by
\begin{align*}
    \beta_0&=\theta_q(+a_0,+a_1,+a_2),\\
     \beta_1&=\theta_q(-a_0,+a_1,-a_2),\\
      \beta_2&=\theta_q(+a_0,-a_1,-a_2),\\
       \beta_3&=\theta_q(-a_0,-a_1,+a_2). 
\end{align*}
When considering solutions defined on a discrete $q$-spiral, i.e. $t\in q^\mathbb{Z} t_0$, the value of $p:=\rho(t_0)$ uniquely determines the corresponding solution $(f_0,f_1,f_2)$ of $q\Pfour(a)$ \cite{joshiroffelsenrhp}.

In the following proposition, the action of the symmetries on the monodromy coordinates is determined.
\begin{proposition}\label{prop:actioncoordinate}
The symmetry $\mathcal{T}_+$ acts on the monodromy coordinates by
\begin{equation*}
    \rho_k(t)\mapsto \widetilde{\rho}_k(t)=-\rho_k(1/t)\quad (k=1,2,3).
\end{equation*}
The symmetry $\mathcal{T}_-$ acts on the monodromy coordinates by
\begin{equation*}
    \rho_k(t)\mapsto \widetilde{\rho}_k(t)=-\frac{1}{\rho_k(-1/t)}\quad (k=1,2,3).
\end{equation*}
\end{proposition}
\begin{proof}
To compute the action of the symmetries on the monodromy coordinates, we need some basic facts about the operator $\pi(\cdot)$. Firstly, given any rank one $2\times 2$ matrix $R$, and  invertible $2\times 2$ matrix $N=(n_{ij})$, we have
\begin{equation}\label{eq:pi_identities}
    \pi(RN)=\pi(R),\quad \pi(NR)=\chi_N(\pi(R)),
\end{equation}
where $\chi_N$ denotes the m\"obius transformation
\begin{equation*}
    \chi_N(z)=\frac{n_{11}z+n_{12}}{n_{21}z+n_{22}}.
\end{equation*}
In particular,
\begin{equation*}
    \pi(\sigma_1 R)=\chi_{\sigma_1}(\pi(R))=1/\pi(R).
\end{equation*}

Secondly, it is elementary to check that
\begin{equation*}
    \pi(R^{\diamond})=-1/\pi(R).
\end{equation*}

We now compute, for transformation $\mathcal{T}_+$,
\begin{align*}
    \widetilde{\rho}_k(t)&=\pi[\widetilde{C}(x_k,t)]=\pi[i\,\sigma_1C^{\diamond}(x_k,1/t)]=\pi[\sigma_1C^{\diamond}(x_k,1/t)]\\
    &=1/\pi[C^{\diamond}(x_k,1/t)]=-\pi[C(x_k,1/t)]=-\rho_k(1/t).
\end{align*}

Similarly, for transformation $\mathcal{T}_-$, we have
\begin{align*}
    \widetilde{\rho}_k(t)&=\pi[\widetilde{C}(x_k,t)]=\pi[-i\,C^{\diamond}(x_k,-1/t)\sigma_3]=\pi[C^{\diamond}(x_k,-1/t)]\\
    &=-\frac{1}{\pi[C(x_k,-1/t)]}=-\frac{1}{\rho_k(-1/t)},
\end{align*}
and the proposition follows.
\end{proof}

In the sequel, the following technical lemma will be of importance. Its proof is given in Appendix \ref{app:technical_lemma}.
\begin{lemma}\label{lem:forbidden}
Let $t_0$, with $t_0^2\notin q^{\mathbb{Z}}$, be inside the domain of a solution $f=(f_0,f_1,f_2)$ of $q\Pfour$. If $f(t)$ takes at least one non-singular value, i.e. a value in $(\mathbb{C}^*)^3$, at a point $t\in q^\mathbb{Z}t_0$, then the coordinates $p=\rho(t_0)$ cannot lie on the curve defined by the intersection of the following equations in $(\mathbb{CP}^1)^3$,
\begin{align}\label{eq:forbidden_cubic}
0=&+\beta_0 p_1p_2p_3-\beta_1 p_1+\beta_2 p_2-\beta_3 p_3,\\
0=&+\beta_0-\beta_1 p_2p_3+\beta_2 p_1p_3-\beta_3 p_1p_2,\nonumber
\end{align}
with the same coefficients as the cubic \eqref{eq:cubic}.
We note that points on this curve solve the cubic equation \eqref{eq:cubic} irrespective of the value of $t$.
\end{lemma}

Let us now take any solution $f$ of $q\Pfour(a)|_{t_0=i}$ on the $q$-spiral $q^\mathbb{Z}i$. To it, corresponds a unique triplet $p=(p_1,p_2,p_3)$, defined by $p_k:=\rho_k(i)$, $k=1,2,3$, which satisfies the cubic equation
\begin{align*}
0=&+\theta_q(+a_0,+a_1,+a_2)\left(p_1p_2p_3-i\right)\\
&-\theta_q(-a_0,+a_1,-a_2)\left(p_1-i\,p_2p_3\right)\\
&+\theta_q(+a_0,-a_1,-a_2)\left(p_2-i\,p_1p_3\right)\\
&-\theta_q(-a_0,-a_1,+a_2)\left(p_3-i\,p_1p_2\right),
\end{align*}
as follows from the identity $\theta_q(-i)=i\,\theta_q(i)$, and does not lie on the curve defined by by equations \eqref{eq:forbidden_cubic}.

Note that $\widetilde{f}=\mathcal{T}_-(f)$ defines another solution on the same domain $q^\mathbb{Z}i$, and its monodromy coordinates, $\widetilde{p}_k:=\widetilde{\rho}_k(i)$, $k=1,2,3$, are related to those of $f$ by
\begin{equation*}
    \widetilde{p}_k=-1/p_k\quad (k=1,2,3).
\end{equation*}
In particular, $f$ is a symmetric solution if and only if $\widetilde{f}=f$, which in turn is equivalent to
\begin{equation}\label{eq:coordinate_symmetric}
    p_k=-1/p_k\quad (k=1,2,3).
\end{equation}
In other words, symmetric solutions of $q\Pfour(a)|_{t_0=i}$ correspond to monodromy coordinates $p$ which satisfy the cubic equation above as well as \eqref{eq:coordinate_symmetric}.

We proceed to compute four triples $p$ that satisfy these conditions. Firstly, equation \eqref{eq:coordinate_symmetric} has only two solutions in $\mathbb{CP}^1$, given by $\pm i$, and we may thus set $p_k=\epsilon_k i$, $\epsilon_k=\pm 1$, $k=1,2,3$. Substitution of these into the cubic shows that the latter is identically zero if the epsilons satisfy
\begin{equation*}
    \epsilon_1\epsilon_2\epsilon_3=-1,
\end{equation*}
as in such a case
\begin{equation*}
    p_1p_2p_3-i=p_j-i\,p_kp_l=0\qquad (\{j,k,l\}=\{1,2,3\}).
\end{equation*}
In particular, this gives us four solutions,
\begin{equation}\label{eq:rhopossible}
    (p_1,p_2,p_3)\in\{(-i,-i,-i), (-i,i,i),(i,-i,i),(i,i,-i)\},
\end{equation}
corresponding to the four  symmetric solutions in Lemma \ref{lem:symmetric_solutions}.

Whilst for generic values of the parameters, these are the only solutions to the cubic, it may so happen for special values of the parameters, that there is a choice of epsilons, with
\begin{equation*}
    \epsilon_1\epsilon_2\epsilon_3=+1,
\end{equation*}
that also solves the cubic. But in such a case, a direct computation yields
\begin{equation*}
    -\beta_0-\beta_1 \epsilon_1+\beta_2 \epsilon_2-\beta_3\epsilon_3=0,
\end{equation*}
and thus the point $(p_1,p_2,p_3)$ lies on the curve \eqref{eq:forbidden_cubic} and hence does not correspond to a solution of $q\Pfour$.

In the next section, Section \ref{sec:solve_linear}, we derive which values of the coordinates in equation \eqref{eq:rhopossible} correspond to which initial conditions
\begin{equation*}
   (f_0(i),f_1(i),f_2(i))\in \{(-1,-1,-1),(-1,1,1),(1,-1,1),(1,1,-1)\}.
\end{equation*}
We answer this question by explicitly solving the linear problem at the reflection point $t=i$ for each case; see Theorem \ref{thm:solvable}.

%% file: monodromy_solvable.tex
\section{Explicit solvability of the linear problem at a reflection point}\label{sec:solve_linear}
In this section we show that the linear problem is explicitly solvable at the reflection point $t_0=i$, for symmetric solutions. In particular, we will prove the following theorem in the end of Section \ref{subsec:construct_global}.

\begin{theorem}\label{thm:solvable}
Let $(f_0,f_1,f_2)$ be a symmetric solution of $q\Pfour(a)|_{t_0=i}$, invariant under $\mathcal{T}_-$, satisfying initial conditions
\begin{equation*}
    (f_0(i),f_1(i),f_2(i))=(v_0,v_1,v_2),
\end{equation*}
so that (by Lemma \ref{lem:symmetric_solutions}),
\begin{equation*}
    (v_0,v_1,v_2)\in \{(-1,-1,-1),(-1,1,1),(1,-1,1),(1,1,-1)\}.
\end{equation*}
Fix the auxiliary functions $u$ and $d$ by the initial conditions $u(i)=1$ and $d(i)=i$. Then, the connection matrix at $t=i$ is explicitly given by
\begin{equation}\label{eq:connection_explicit}
C(z,i)=2 c_0^3\begin{pmatrix}
h(i\,z) & i\, h(-i\,z)\\
-h(-i\,z) & i\, h(i\,z)
\end{pmatrix},    
\end{equation}
where the scalar $c_0$ equals
\begin{equation*}
    c_0=\frac{\sqrt{i}\,\theta_q(i)}{\sqrt{2}\,\theta_q(-1)}=\frac{1}{2}\prod_{k=1}^\infty{ \frac{(1+q^ki)(1-q^ki)}{(1+q^k)^2}},
\end{equation*}
and the function $h(z)$ is defined by
\begin{align*}
    h(z)=&+\theta_q\bigg(+\frac{v_1}{x_1}z,-\frac{v_0}{x_2}z,+\frac{v_2}{x_3} z\bigg)-\theta_q\bigg(+\frac{v_1}{x_1}z,+\frac{v_0}{x_2}z,-\frac{v_2}{x_3} z\bigg)\\
    &-\theta_q\bigg(-\frac{v_1}{x_1}z,-\frac{v_0}{x_2}z,-\frac{v_2}{x_3} z\bigg)-\theta_q\bigg(-\frac{v_1}{x_1}z,+\frac{v_0}{x_2}z,+\frac{v_2}{x_3} z\bigg),
\end{align*}
with
\begin{equation*}
    (x_1,x_2,x_3)=(a_0^{-1},a_1/q,q^{-1}).
\end{equation*}
In particular, the corresponding values of the monodromy coordinates, $p_k=\rho_k(i)$, $k=1,2,3$, are given by
\begin{equation}\label{eq:p_explicit}
    (p_1,p_2,p_3)=(-v_1i,v_0i,-v_2i).
\end{equation}
\end{theorem}
\begin{remark}
In the proof of Theorem \ref{thm:solvable}, we also obtain the following alternative
expression for the connection matrix,
\begin{equation*}
    C(z)=\sigma_1 C_0\left(\frac{v_1}{x_1} z\right) M
    C_0\left(-\frac{v_0}{x_2} z\right)MC_0\left(\frac{v_2}{x_3} z\right),
\end{equation*}
where $C_0(z)$, given in Proposition \ref{prop:connection}, is the connection matrix of a degree one Fuchsian system and the matrix $M$ is defined in equation \eqref{eq:defiM}.
\end{remark}

The spectral equation of the Lax pair \eqref{eq:laxpair} naturally comes in a factorised form. The fundamental reason that allows us to solve the linear problem at the reflection point $t=i$, for a symmetric solution as in Theorem \ref{thm:solvable}, is that the factors in this form `almost' commute. Namely, by fixing $u(i)=1$, we have
\begin{equation*}
    A(z,i)=A_0\left(\frac{v_2z}{x_3}\right)
    A_0\left(\frac{v_0z}{x_2}\right)
    A_0\left(\frac{v_1z}{x_1}\right),
\end{equation*}
where 
\begin{equation*}
    A_0(z)=i\,\sigma_2+z\,\sigma_3,
\end{equation*}
and these factors satisfy the commutation relation,
\begin{equation}\label{eq:commutation_relation}
    A_0(x)A_0(y)=A_0(-y)A_0(-x).
\end{equation}
This observation allows us to construct global solutions of the linear system
\begin{equation*}
    Y(qz)=A(z,i)Y(z),
\end{equation*}
from solutions of the simpler system
\begin{equation*}
    U(qz)=A_0(z)U(z),
\end{equation*}
which we will refer to as the model problem.

In Section \ref{subsec:model_solve}, we solve this model problem, and in Section \ref{subsec:construct_global} we use this to construct global solutions of the spectral equation at $t=i$ and prove Theorem \ref{thm:solvable}. The model problem is solved in terms of basic hypergeometric functions, denoted for given parameter $a$, $0<p<1$ and $z\in\mathbb C$ by 
\[
_{0}\phi_1\left[\begin{matrix} 
\text{-} \\ 
a \end{matrix} 
; p, z \right],
\]
whose mathematical properties can be found in \cite{gasper}.

\subsection{The model problem}\label{subsec:model_solve}
In this section, we study the model problem,
\begin{equation*}
    U(qz)=A_0(z)U(z),\quad A_0(z)=i\,\sigma_2+z\sigma_3.
\end{equation*}
Firstly, we find an explicit expression for the canonical solution at $z=\infty$. 
\begin{lemma}\label{lem:model_sol_inf}
There exists a unique matrix function $U_\infty(z)$, analytic on $\mathbb{C}^*$, which solves
\begin{equation}\label{eq:power_series}
    U_\infty(qz)=z^{-1}A_0(z)U_\infty(z)\sigma_3,\qquad U_\infty(z)=I+\mathcal{O}(z^{-1})\quad (z\rightarrow \infty),
\end{equation}
explicitly  given by
\begin{equation*}
    U_\infty(z)=g_\infty(z)I+h_\infty(z)\sigma_1,
\end{equation*}
where $g_\infty(z)$ and $h_\infty(z)$ are the basic hypergeometric functions,
\begin{align*}
g_\infty(z)&=\;_{0}\phi_1 \left[\begin{matrix} 
\text{-} \\ 
-q \end{matrix} 
; q^2,-\frac{q^3}{z^2} \right],\\
    h_\infty(z)&=-\frac{q}{(q+1)z}\;_{0}\phi_1 \left[\begin{matrix} 
\text{-} \\ 
-q^3 \end{matrix} 
; q^2,-\frac{q^5}{z^2} \right].
\end{align*}
\end{lemma}
\begin{proof}
It is an elementary computation to show that \eqref{eq:power_series} has a unique formal power series solution around $z=\infty$. Furthermore, by using the defining formula,
\begin{equation}\label{eq:hyper}
    \;_{0}\phi_1 \left[\begin{matrix} 
\text{-} \\ 
b\end{matrix} 
; p,x \right]=\sum_{n=0}^\infty \frac{p^{n(n-1)}}{(b;p)_n(p;p)_n}x^n,
\end{equation}
it is checked directly that this formal power series solution is indeed given by $U_\infty(z)$. Since, furthermore, the series \eqref{eq:hyper} has infinite radius of convergence, $U_\infty(z)$ is an analytic function on $\mathbb{CP}^1\setminus\{0\}$, which thus uniquely solves equation \eqref{eq:power_series}, and the lemma follows. 
\end{proof}

We have a similar result near $z=0$.

\begin{lemma}
Define
\begin{equation}\label{eq:defiM}
    M=\begin{pmatrix}
    1 & -1\\
    i & i
    \end{pmatrix},
\end{equation}
so that $M^{-1}(i\,\sigma_2)M=i\,\sigma_3$. Then, there exists a unique matrix function $U_0(z)$, meromorphic on $\mathbb{C}$, which satisfies
\begin{equation*}
U_0(qz)=A_0(z)U_0(z)(i\,\sigma_3)^{-1},\qquad U_0(z)=M+\mathcal{O}(z)\quad (z\rightarrow 0),
\end{equation*}
explicitly given by
\begin{equation*}
    U_0(z)=\frac{1}{(+z;q)_\infty(-z;q)_\infty}M\cdot (g_0(z)I+h_0(z)\sigma_2),
\end{equation*}
where
\begin{align*}
g_0(z)&=\;_{0}\phi_1 \left[\begin{matrix} 
\text{-} \\ 
-q \end{matrix} 
; q^2,- q z^2\right],\\
    h_0(z)&=\frac{z}{q+1}\;_{0}\phi_1 \left[\begin{matrix} 
\text{-} \\ 
-q^3 \end{matrix} 
; q^2,-q^3 z^2 \right].
\end{align*}
\end{lemma}
\begin{proof}
This is proven analogously to Lemma \ref{lem:model_sol_inf}.
\end{proof}

In the following proposition, we explicitly determine the connection matrix of the model problem.
\begin{proposition}\label{prop:connection}
The connection matrix
\begin{equation}\label{eq:connection}
    C_0(z)=U_0(z)^{-1}U_\infty(z),
\end{equation}
is given by
\begin{equation*}
    C_0(z)=c_0\left(\theta_q(+i z)\begin{pmatrix}
    1 & 0\\
    0 & -i
    \end{pmatrix}+\theta_q(-i z)\begin{pmatrix}
    0 & -i\\
    -1 & 0
    \end{pmatrix}\right),
\end{equation*}
where the scalar $c_0$ is given by
\begin{equation*}
    c_0:=\frac{\sqrt{i}\,\theta_q(i)}{\sqrt{2}\,\theta_q(-1)}=\frac{1}{2}\prod_{k=1}^\infty{ \frac{(1+q^ki)(1-q^ki)}{(1+q^k)^2}}.
\end{equation*}    
\end{proposition}
\begin{proof}
From the defining properties of $U_\infty(z)$ and $U_0(z)$, it follows that
\begin{equation}\label{eq:sol_det}
    |U_\infty(z)|=(+z,q)_\infty (-z,q)_\infty,\quad |U_0(z)|=2 i(+q/z,q)_\infty^{-1} (-q/z,q)_\infty^{-1}.
\end{equation}
In particular, $C_0(z)$ is an analytic function on $\mathbb{C}^*$. Furthermore, it satisfies
\begin{equation*}
    C_0(qz)=i\,z^{-1}\sigma_3C_0(z)\sigma_3,
\end{equation*}
and its entries are thus degree one $q$-theta functions, i.e.
\begin{equation*}
    C_0(z)=\theta_q(+i z)\begin{pmatrix}
    c_{11} & 0\\
    0 & c_{22}
    \end{pmatrix}+\theta_q(-i z)\begin{pmatrix}
    0 & c_{12}\\
    c_{21} & 0
    \end{pmatrix},
\end{equation*}
for some $c_{ij}\in\mathbb{C}, 1\leq i,j\leq 2$.

Now, observe that
\begin{equation*}
    U_\infty(z)^{\diamond}=\sigma_3 U_\infty(z)\sigma_3,\quad U_0(z)^{\diamond}=i\,\sigma_3 U_0(z),
\end{equation*}
and therefore
\begin{equation*}
    C_0(z)^{\diamond}=-i\,C_0(z)\sigma_3.
\end{equation*}
We thus find the following conditions on the coefficients,
\begin{equation*}
    c_{11}=i\,c_{22},\quad c_{12}=i\,c_{21}.
\end{equation*}

Due to equations \eqref{eq:sol_det}, we have
\begin{equation*}
    |C_0(z)|=\frac{1}{2i}\theta_q(+z)\theta_q(-z).
\end{equation*}
Evaluating this identity at $z=i$, gives
\begin{equation*}
 -i\theta_q(-1)^2 c_{11}^2=\frac{1}{2i}\theta_q(+i)\theta_q(-i)=\frac{1}{2}\theta_q(i)^2,
\end{equation*}
and therefore $c_{11}^2=c_0^2$. Similarly, we obtain $c_{21}^2=c_0^2$, so that
\begin{equation*}
    c_{11}=\epsilon_1 c_0,\quad c_{21}=\epsilon_2 c_0,
\end{equation*}
for some $\epsilon_{1,2}\in\{\pm 1\}$.

Note that $\epsilon_{1,2}$ must be continuous functions of $q$ in the punctured unit disc $\{0<|q|<1\}$ and they are thus global constants. We now choose $0<q<1$, so that
\begin{equation*}
    \overline{U_\infty(\overline{z})}=U_\infty(z),\quad \overline{U_0(\overline{z})}=-U_0(z)\sigma_1.
\end{equation*}
In particular, this means that
\begin{equation*}
    \overline{C_0(\overline{z})}=-\sigma_1 C_0(z),
\end{equation*}
and, by noting that $\overline{c_0}=c_0$, we thus obtain $\epsilon_1=\epsilon_2$.

It only remains to be checked that $\epsilon_1=1$. To this end, note that equation \eqref{eq:connection} implies the following connection result,
\begin{equation*}
    g_\infty(z)=\frac{\epsilon_1 c_0}{(z^2,q^2)_\infty} \left[(\theta_q(i\, z)+\theta_q(-i\,z))g_0(z)-i(\theta_q(i\, z)-\theta_q(-i\,z))h_0(z)\right].
\end{equation*}
Setting $z=i\,x$, with $0<x<\infty$, we thus have
\begin{equation}\label{eq:positive_identity}
    g_\infty(i\,x)=\frac{\epsilon_1 c_0}{(-x^2,q^2)_\infty} \left[(\theta_q(-x)+\theta_q(x))g_0(z)+(\theta_q(-x)-\theta_q(x))(-i\,h_0(i\, x))\right].
\end{equation}
We claim that each of the terms
\begin{equation*}
    g_\infty(i\,x),\quad g_0(i\,x),\quad -i \,h_0(i\,x),\quad (-x^2,q)_\infty,\quad \theta_q(-x)\pm\theta_q(x),
\end{equation*}
 is a real and positive function of $x$ on $(0,+\infty)$. For example, the inequality $(-x;q)_\infty>(+x;q)_\infty$, on the positive real line, follows almost directly from the definition of the $q$-Pochammer symbol,  
 and thus
 \begin{equation*}
     b(x):=\theta_q(-x)-\theta_q(+x)>0,
 \end{equation*}
 on the positive real line. Therefore, also
  \begin{equation*}
  \theta_q(-x)+\theta_q(+x)=x\,b(q\,x)>0,
 \end{equation*}
 on the positive real line. Each of the hypergeometric series, $g_\infty(i\,x), g_0(i\,x), -i \,h_0(i\,x)>0$, on the positive real line, since all the coefficients in the different series are positive.

Since $c_0>0$, equation \eqref{eq:positive_identity} can thus only hold if $\epsilon_1=+1$, and the proposition follows.
\end{proof}

\begin{corollary}\label{cor:connection} The explicit expression for the connection matrix in Proposition \ref{prop:connection}, yields the following connection formulas,
\begin{align*}
    \;_{0}\phi_1 \left[\begin{matrix} 
\text{-} \\ 
-q \end{matrix} 
; q^2,-\frac{q^3}{z^2} \right]=&+
\frac{c_0(\theta_q(-i\, z)+\theta_q(i\, z))}{(z^2;q^2)_\infty}\;_{0}\phi_1 \left[\begin{matrix} 
\text{-} \\ 
-q \end{matrix} 
; q^2,-q z^2\right]\\
&+
\frac{c_0\, i\, z(\theta_q(-i\, z)-\theta_q(i\, z))}{(1+q)(z^2;q^2)_\infty}\;_{0}\phi_1 \left[\begin{matrix} 
\text{-} \\ 
-q^3 \end{matrix} 
; q^2,-q^3 z^2 \right],
\\
\;_{0}\phi_1 \left[\begin{matrix} 
\text{-} \\ 
-q^3 \end{matrix} 
; q^2,-\frac{q^5}{z^2} \right]=&+
\frac{(1+q)c_0\, i\,z(\theta_q(-i\, z)-\theta_q(i\, z))}{q(z^2;q^2)_\infty}\;_{0}\phi_1 \left[\begin{matrix} 
\text{-} \\ 
-q \end{matrix} 
; q^2,-q z^2\right]\\
&+
\frac{c_0\,z^2(\theta_q(-i\, z)+\theta_q(i\, z))}{q(z^2;q^2)_\infty}\;_{0}\phi_1 \left[\begin{matrix} 
\text{-} \\ 
-q^3 \end{matrix} 
; q^2,-q^3 z^2 \right],
\end{align*}
where the value of $c_0$ is given in Proposition \ref{prop:connection}.
\end{corollary}

\begin{remark}
Note that the solutions to the model problem are essentially built out of Jackson's $q$-Bessel functions of the second kind,
\begin{equation*}
    J_\nu^{(2)}(x;p)=\frac{(p^{\nu+1};p)_\infty}{(p;p)_\infty}\left(\frac{x}{2}\right)^\nu 
    \;_{0}\phi_1 \left[\begin{matrix} 
\text{-} \\ 
p^{\nu+1} \end{matrix} 
; p,-\frac{x^2 p^{\nu+1}}{4} \right],
\end{equation*}
with $p=q^2$ and $\nu=\pm\tfrac{1}{2}$. In particular, we could have alternatively used the known connection results for these functions \cites{zhangbessel,moritabessel}, in conjunction with transformation formulas for $\;_{0}\phi_1$ hypergeometric functions \cite{gasper},
to obtain the connection formulas in Corollary \ref{cor:connection} and, consequently,
Proposition \ref{prop:connection}.
\end{remark}

\subsection{Constructing global solutions}\label{subsec:construct_global}
In this section, we construct solutions of the spectral equation at $t=i$ given by
\begin{equation*}
    Y(qz)=A(z,i)Y(z),\quad 
    A(z,i)=A_0\left(\frac{v_2z}{x_3}\right)
    A_0\left(\frac{v_0z}{x_2}\right)
    A_0\left(\frac{v_1z}{x_1}\right).
\end{equation*}
Motivated by the commutation relation \eqref{eq:commutation_relation}, we consider the ansatz
\begin{equation}\label{eq:phiinf}
    \Phi_\infty(z)=U_\infty(r_1 z)U_\infty(r_2 z)U_\infty(r_3 z),
\end{equation}
for the matrix function $\Phi_\infty(z)$ defined in Lemma \ref{lem:solinf}, for some $r_1,r_2,r_3$ to be determined. Using the commutation relation
\begin{equation*}
    U_\infty(xz)\sigma_3A_0(yz)=\sigma_3A_0(yz)U_\infty(xz),
\end{equation*}
we find
\begin{align*}
    \Phi_\infty(qz)&=U_\infty(qr_1 z)U_\infty(qr_2 z)U_\infty(qr_3 z),\\
    &=\frac{1}{r_1r_2r_3 z^3}A_0(r_1z)U_\infty(r_1z)\sigma_3 A_0(r_2z)U_\infty(r_2z)\sigma_3
    A_0(r_3z)U_\infty(r_3z)\sigma_3\\
    &=\frac{1}{r_1r_2r_3 z^3}A_0(r_1z)\sigma_3 A_0(r_2z)U_\infty(r_1z)\sigma_3
    A_0(r_3z)U_\infty(r_2z)U_\infty(r_3z)\sigma_3\\
    &=\frac{1}{r_1r_2r_3 z^3}A_0(r_1z)\sigma_3 A_0(r_2z)\sigma_3
    A_0(r_3z)U_\infty(r_1z)U_\infty(r_2z)U_\infty(r_3z)\sigma_3\\
    &=\frac{1}{r_1r_2r_3 i z^3}A_0(r_1z)A_0(-r_2z)
    A_0(r_3z)U_\infty(r_1z)U_\infty(r_2z)U_\infty(r_3z)(i\sigma_3)^{-1}.
\end{align*}
Therefore, if we set
\begin{equation}\label{eq:defi_r}
    (r_1,r_2,r_3)=\left(\frac{v_2}{x_3},-\frac{v_0}{x_2},\frac{v_1}{x_1}\right),
\end{equation}
then $\Phi_\infty(z)$ solves
\begin{equation*}
   \Phi_\infty(qz)= \frac{1}{qa_0^2a_2i}z^{-3}A(z,i)\Phi_\infty(z)(i\sigma_3)^{-1}.
\end{equation*}
Furthermore, note that $\Phi_\infty(z)=I+\mathcal{O}(z^{-1})$ as $z\rightarrow \infty$, so that our ansatz is indeed correct for the choice of $(r_1,r_2,r_3)$ above.

Similarly, using the commutation relation
\begin{equation*}
    U_0(xz)M^{-1}(i\,\sigma_2)A_0(yz)=(i\,\sigma_2)A_0(yz)U_0(xz)M^{-1},
\end{equation*}
it follows that
\begin{equation}\label{eq:phi0}
    \Phi_0(z)=U_0(r_1z)M^{-1}U_0(r_2z)M^{-1}U_0(r_3z)\sigma_1,
\end{equation}
satisfies
\begin{align*}
   \Phi_0(qz)&=A(z,i)\Phi_0(z)(i\sigma_3)^{-1},\\
   \Phi_0(z)&=M\sigma_1+\mathcal{O}(z)\quad (z\rightarrow 0),
\end{align*}
for the same choice of $(r_1,r_2,r_3)$. Furthermore, note that
\begin{equation*}
    M\sigma_1=M_0,
\end{equation*}
if we choose $d(i)=i$ in equation \eqref{eq:A0iv}. Therefore, the formula for $\Phi_0(z)$ above is an explicit expression for the canonical matrix function at $z=0$ defined in Lemma \ref{lem:solzero}.

We are now in a position to prove Theorem \ref{thm:solvable}.
\begin{proof}[Proof of Theorem \ref{thm:solvable}]
By definition, the connection matrix at $t=i$ is given by
\begin{equation*}
    C(z,i)=\Phi_0(z)^{-1}\Phi_\infty(z),
\end{equation*}
where $\Phi_\infty(z)$ and $\Phi_0(z)$ are given by the explicit formulas \eqref{eq:phiinf} and \eqref{eq:phi0}. This yields, 
\begin{equation*}
    C(z)=\sigma_1 C_0(r_3 z)U_\infty(r_3 z)^{-1} M
    C_0(r_2 z)U_\infty(r_2 z)^{-1} M
    C_0(r_1 z)U_\infty(r_2 z) U_\infty(r_3 z),
\end{equation*}
where the constants $(r_1,r_2,r_3)$ are defined in equation \eqref{eq:defi_r} and $M$ is defined in equation \eqref{eq:defiM}.

In order to simplify this expression, we use the following commutation relations,
\begin{equation*}
    M \sigma_2=-\sigma_1 M,\quad C_0(z) \sigma_1=-\sigma_2 C_0(z),
\end{equation*}
so that,
\begin{align*}
    MC_0(r_1 z)U_\infty(r_2 z)&=M
    C_0(r_1 z)(g(r_2 z)I+h(r_2 z)\sigma_1)\\
    &=M(g(r_2 z)I-h(r_2 z)\sigma_2)C_0(r_1 z)\\
    &=(g(r_2 z)I+h(r_2 z)\sigma_1)MC_0(r_1 z)\\
    &=U_\infty(r_2 z)MC_0(r_1 z).
\end{align*}
In other words, $MC_0(r_1 z)$ and $U_\infty(r_2 z)$ commute and we thus obtain the following simpler expression for $C(z)$,
\begin{equation*}
    C(z)=\sigma_1 C_0(r_3 z)U_\infty(r_3 z)^{-1} M
    C_0(r_2 z)MC_0(r_1 z) U_\infty(r_3 z).
\end{equation*}
It follows from the computation before, that $MC_0(r_{1,2} z)$ also commutes with $U_\infty(r_3 z)$, and we thus obtain
\begin{equation}\label{eq:Cfactorised}
    C(z)=\sigma_1 C_0(r_3 z) M
    C_0(r_2 z)MC_0(r_1 z).
\end{equation}
It is now a direct computation that yields the explicit expression \eqref{eq:connection_explicit} for $C(z)$.

The same holds true for the expressions for the monodromy coordinates \eqref{eq:p_explicit}, using equation \eqref{eq:connection_explicit}. Rather than going through these computations, we finish the proof of the theorem with an alternative method to compute e.g. $p_1$. Using the factorisation \eqref{eq:Cfactorised}, we find
\begin{align*}
    p_1&=\pi\left[C(x_1)\right]\\
    &=\pi\left[\sigma_1 C_0(r_3 x_1) M
    C_0(r_2 x_1)MC_0(r_1 x_1)\right]\\
    &=\pi\left[\sigma_1 C_0(v_1) M
    C_0(-v_0 x_1/x_2)MC_0(v_2 x_1/x_3)\right].
\end{align*}
Due to the non-resonance conditions \eqref{eq:conditionnonresonant}, neither $|C_0(-v_0 x_1/x_2)|$ nor $|C_0(v_2 x_1/x_3)|$ vanishes, so by identities \eqref{eq:pi_identities} for the $\pi(\cdot)$ operator, we obtain
\begin{align*}
    p_1&=\pi\left[\sigma_1 C_0(v_1)\right]=1/\pi\left[C_0(v_1)\right]=-\frac{\theta_q(-i\, v_1)}{\theta_q(+i\, v_1)}=-\frac{\theta_q(q\,i\, v_1)}{\theta_q(+i\, v_1)}=-i\, v_1.
\end{align*}
Similar computations can be carried out of $p_{2,3}$ and the theorem follows.
\end{proof}

%% file: meromorphic_sol.tex
\section{The monodromy problem of the $q$-Okamoto rational solutions}\label{s:mero}
In this section we consider symmetric solutions of $q\Pfour$ defined on (connected) open subsets of the complex plane. A particular class of such solutions is given by the $q$-Okamoto rational solutions. We study them in detail and show that their monodromy problems are solvable for all values of the independent variable. 

Let $T$ be a non-empty, open and connected subset of the universal covering of $\mathbb{C}^*$, with $qT=T$. We call
a triplet $f=(f_0,f_1,f_2)$ of meromorphic functions on $T$ that satisfies $q\Pfour$ identically, a meromorphic solution of $q\Pfour$. We call it symmetric, when the solution (and its domain) are invariant under $\mathcal{T}_+$ or $\mathcal{T}_-$.

Each meromorphic solution corresponds to a unique triplet $\rho=(\rho_1,\rho_2,\rho_3)$ of complex functions on $T$ that solve the cubic equation \eqref{eq:cubic}
identically in $t$ and the $q$-difference equations
\begin{equation}\label{eq:rho_qdif}
    \rho_k(qt)=-\rho_k(t),\quad (k=1,2,3),
\end{equation}
which follow from the time-evolution of the connection matrix $C(z,t)$ (see equation \eqref{eq:connection_evolution}).

Now, it might happen that, for special values of $t_0\in T$, the value of $f(t)$ does not lie in $(\mathbb{C}^*)^3$, for every $t\in q^\mathbb{Z}t_0$. At such times $t=t_0$, the monodromy coordinates $\rho(t)$ either have an essential singularity, or they lie on the curve defined by equations \eqref{eq:forbidden_cubic}. On the other hand, if $f(t)$ is regular for at least one value of $t\in q^\mathbb{Z}t_0$, then the value of the monodromy coordinates $\rho(t)$ at $t=t_0$ is well-defined and does not lie on the curve given by equations \eqref{eq:forbidden_cubic}.

In the following, we restrict our discussion to considering meromorphic solutions which do not have $q$-spirals of poles. If such a solution is symmetric with respect to $\mathcal{T}_-$, that is,
\begin{equation*}
    f_k(t)=1/f_k(-1/t)\quad (k=0,1,2),
\end{equation*}
then, by Proposition \ref{prop:actioncoordinate}, the $\rho$-coordinates have the same symmetry,
\begin{equation}\label{eq:rho_sym}
\rho_k(t)=-\frac{1}{\rho_k(-1/t)}\quad (k=1,2,3).
\end{equation}
This means that we can classify symmetric meromorphic solutions, in terms of 
 meromorphic triplets $\rho=\rho(t)$ which solve the cubic \eqref{eq:cubic}, as well as equations \eqref{eq:rho_qdif} and \eqref{eq:rho_sym}, and do not hit the curve defined by equations \eqref{eq:forbidden_cubic}.
Similar statements follow for solutions symmetric with respect to $\mathcal{T}_+$, in which case we have
\begin{equation}\label{eq:rho_sym_plus}
\rho_k(t)=-\rho_k(-1/t)\quad (k=1,2,3).
\end{equation}
%If we assume that the domain $T$ contains $t=i$, then there are only four possible values of $\rho(t)$ at $t=i$, given in equation \eqref{eq:rhopossible}. We pick one of them,
%\begin{equation*}
 %   \rho_1(i)=\rho_2(i)=\rho_3(i)=-i.
%\end{equation*}
%The general solution of equation \eqref{eq:rho_sym}, with these initial conditions, is given by
%\begin{equation*}
%    \rho_k(t)=-t\frac{g_k(-1/t)}{g_k(t)},
%\end{equation*}
%where $g_k(t)$ any meromorphic function on $T$, with $g_k(0)=1$, for $k=1,2,3$.

%It is now easy to see that there are many solutions of \eqref{eq:rho_qdif} and \eqref{eq:cubic}, that can be built with appropriate choices of $(g_1,g_2,g_3)$. As a consequence, there exist many different meromorphic extensions of the discrete symmetric solutions found in Lemma \ref{lem:symmetric_solutions}.

In the remainder of this section, we focus on a particular collection of symmetric meromorphic solutions for which we compute the monodromy. These solutions are the $q$-Okamoto rational solutions, which are rational in $t^{\frac{1}{3}}$, derived by Kajiwara et al. \cite{kajiwaranoumiyamada2001}.
\begin{theorem}[Kajiwara et al. \cite{kajiwaranoumiyamada2001}] \label{thm:oka_rational}
For $m,n\in\mathbb{Z}$, the formulas
\begin{align*}
    f_0&=x^2 r^{2n-m}\frac{Q_{m+1,n}(r^{+1}x^2)Q_{m+1,n+1}(r^{-1} x^2)}{Q_{m+1,n}(r^{-1}x^2)Q_{m+1,n+1}(r^{+1} x^2)},\\
    f_1&=x^2 r^{-m-n}\frac{Q_{m+1,n+1}(r^{+1} x^2)Q_{m,n}(r^{-1}x^2)}{Q_{m+1,n+1}(r^{-1}x^2)Q_{m,n}(r^{+1} x^2)},\\
    f_2&=x^2 r^{2m-n}\frac{Q_{m,n}(r^{+1}x^2)Q_{m+1,n}(r^{-1}x^2)}{Q_{m,n}(r^{-1}x^2)Q_{m+1,n}(r^{+1}x^2)},
\end{align*}
give a solution of $q\Pfour$ rational in $x=t^{\frac{1}{3}}$, with parameters
\begin{equation*}
    a_0=r q^m,\quad a_1=r q^{n-m},\quad a_2=r q^{-n},\quad r:=q^{\frac{1}{3}},
\end{equation*}
in terms of the $q$-Okamoto polynomials $Q_{m,n}(x)$  defined through the recurrence relations
\begin{align}\nonumber
Q_{m-1,n}(x/r)Q_{m+1,n+1}(r\, x)=&Q_{m,n}(x/r)Q_{m,n+1}(r\, x)+\\
&x\,Q_{m,n+1}(x/r)Q_{m,n}(r\, x) r^{2m+2n-1},\nonumber\\
Q_{m+1,n}(x/r)Q_{m,n+1}(r\, x)=&Q_{m+1,n+1}(x/r)Q_{m,n}(r\, x)+\label{eq:okamotorecurrence}\\
&x\, Q_{m,n}(x/r)Q_{m+1,n+1}(r\, x)r^{2n-4m+1},\nonumber\\
Q_{m+1,n+1}(x/r)Q_{m,n-1}(r\, x)=&Q_{m,n}(x/r)Q_{m+1,n}(r\, x)+\nonumber\\
&x\, Q_{m+1,n}(x/r)Q_{m,n}(r\, x)r^{2m-4n+1},\nonumber
\end{align}
with $Q_{0,0}(x)=Q_{1,0}(x)=Q_{1,1}(x)=1$.
\end{theorem}

From the recurrence relations for the $q$-Okamoto polynomials, it follows that $Q_{m,n}(x)$ is a monic polynomial of degree $d_{m,n}:=m^2+n^2-m(n+1)$. Furthermore, it can be shown by induction that the polynomials are palindromic, i.e.
\begin{equation}\label{eq:palindrome}
    x^{d_{m,n}}Q_{m,n}(1/x)=Q_{m,n}(x),
\end{equation}
for $m,n\in\mathbb{Z}$. It follows that, upon writing $f_k=f_k(x)$, the corresponding rational solutions defined in Theorem \ref{thm:oka_rational}, satisfy
\begin{equation*}
    f_k(x)=1/f_k(\pm 1/x),
\end{equation*}
for $0\leq k\leq 2$ and any choice of sign. In other words, they are invariant under both $\mathcal{T}_+$ and $\mathcal{T}_-$.

Now consider the branch of $x=x(t)$ which evaluates to $x=-i$ at $t=i$. There, the $q$-Okamoto rationals specialise to the symmetric solutions on discrete time domains classified in Lemma \ref{lem:symmetric_solutions}. To see this, it is helpful to note that equation \eqref{eq:palindrome} implies
\begin{equation*}
    (-r)^{d_{m,n}}Q_{m,n}(-1/r)=Q_{m,n}(-r).
\end{equation*}
Thus, at $x=-i$, so that $t=i$,
\begin{align*}
    f_0(i)&=-r^{2n-m}\frac{Q_{m+1,n}(-r^{+1})Q_{m+1,n+1}(-r^{-1})}{Q_{m+1,n}(-r^{-1})Q_{m+1,n+1}(-r^{+1})},\\
    &=-r^{2n-m}(-r)^{d_{m+1,n}-d_{m+1,n+1}},\\
    &=(-1)^{1+m}.
\end{align*}
By similar computations for $f_1(i)$ and $f_2(i)$, we obtain
\begin{equation}\label{eq:initiali}
f_0(i)=(-1)^{1+m},\quad
    f_1(i)=(-1)^{1+m+n},\quad 
    f_2(i)=(-1)^{1+n}.
\end{equation}
So depend on the values of $m,n\in\mathbb{Z}$, the $q$-Okamoto rational solutions specialise to the different symmetric solutions in Lemma \ref{lem:symmetric_solutions}, on the $q$-spiral $q^{\mathbb{Z}}i$.

\subsection{Solvable monodromy for the seed solution}
In this section, we consider the simplest member of the family of rational solutions defined in Theorem  \ref{thm:oka_rational}, corresponding to $m=n=0$. The parameters of $q\Pfour$ then read
\begin{equation*}
    a_0=a_1=a_2=r,
\end{equation*}
and
\begin{equation*}
    f_0=f_1=f_2=x^2.
\end{equation*}
We call this solution the seed solution. The corresponding value of $b$ in \eqref{eq:bb} is given by
\begin{equation*}
    b=\frac{i\,x}{1-r x^2},
\end{equation*}
and explicit solutions to the auxiliary equations \eqref{eq:auxiliary} and \eqref{eq:timeevolutiond} are given by
\begin{equation*}
    u(x)=\frac{(r\, x^2; r^2)_\infty^2}{\theta_r(x)^2},\quad d(x)=\frac{\theta_r(-x)}{(r\,x^2;r^2)_\infty}.
\end{equation*}

In this special case, the matrix polynomial in the spectral equation \eqref{eq:laxspectral} factorises as
\begin{equation*}
A(z,x)=\begin{pmatrix}
u & 0\\
0 & 1\end{pmatrix} A_1(r^2z,x)A_1(r\, z,x)A_1(z,x)
\begin{pmatrix}
u^{-1} & 0\\
0 & 1\end{pmatrix},
\end{equation*}
with
\begin{equation*}
    A_1(z,x)=\begin{pmatrix}
        -i \,r \,x\,z & 1\\
        -1 & -i\, r/x\, z
    \end{pmatrix}.
\end{equation*}
This means that any solution of
\begin{equation}\label{eq:simpleqdif}
    Y(rz)=\begin{pmatrix}
u & 0\\
0 & 1\end{pmatrix} A_1(z,x)\begin{pmatrix}
u^{-1} & 0\\
0 & 1\end{pmatrix}Y(z),
\end{equation}
also defines a solution of the spectral equation. A classical result \cite{lecaine} shows that equation \eqref{eq:simpleqdif} can be solved in terms of Heine's $q$-hypergeometric functions. We can thus leverage the connection results by Watson \cite{watson}, see also \cite[Section 4.3]{gasper}, to compute the connection matrix of the spectral equation.

We find that the matrix function $\Phi_\infty$, defined in Lemma \ref{lem:solinf}, is given explicitly by
\begin{align*}
  \Phi_\infty(z,t)&=(z^{-1};r)_\infty\begin{pmatrix}
u & 0\\
0 & 1\end{pmatrix} \widehat{\Phi}_\infty(z,x)\begin{pmatrix}
u^{-1} & 0\\
0 & 1\end{pmatrix},\\
\widehat{\Phi}_\infty(z,x)&=\begin{pmatrix}
\hspace{11mm}\;_{2}\phi_1 \left[\begin{matrix} 
1/x, -1/x \\ 
1/x^2\end{matrix} 
; r,\frac{1}{z} \right] & \frac{i\, x}{(1-r x^2)z}
\;_{2}\phi_1 \left[\begin{matrix} 
r\,x, -r\,x \\ 
r^2 x^2\end{matrix} 
; r,\frac{1}{z} \right] \vspace{1mm}\\
\frac{i\, x}{(r-x^2)z}
\;_{2}\phi_1 \left[\begin{matrix} 
r/x, -r/x \\ 
r^2/x^2\end{matrix} 
; r,\frac{1}{z} \right] &
\hspace{8mm}\;_{2}\phi_1 \left[\begin{matrix} 
x, -x \\ 
x^2\end{matrix} 
; r,\frac{1}{z} \right]\\
\end{pmatrix}.
\end{align*}
The matrix function $\Phi_0$, defined in Lemma \ref{lem:solzero}, is given by
\begin{align*}
  \Phi_0(z,t)&=\frac{d}{(r z;r)_\infty}\begin{pmatrix}
u & 0\\
0 & 1\end{pmatrix} \widehat{\Phi}_0(z,x),\\
\widehat{\Phi}_0(z,x)&=\begin{pmatrix}
\hspace{1.1mm}i\;_{2}\phi_1 \left[\begin{matrix} 
-1/x, -r \,x \\ 
-r\end{matrix} 
; r,-r\, z \right] & -i
\;_{2}\phi_1 \left[\begin{matrix} 
1/x, r\,x \\ 
-r\end{matrix} 
; r,-r\, z\right] \vspace{1mm}\\
\;_{2}\phi_1 \left[\begin{matrix} 
-x, -r/x \\ 
-r\end{matrix} 
; r,-r\, z \right] &
 \hspace{4.3mm}\;_{2}\phi_1 \left[\begin{matrix} 
x, -r/x \\ 
-r\end{matrix} 
; r,-r\, z\right]\\
\end{pmatrix}.
\end{align*}
The corresponding connection matrix is then 
\begin{align*}
    C(z,t)&=\widetilde{C}(z,x)\begin{pmatrix}
        d^{-1}u^{-1} & 0\\
        0 & d^{-1}
    \end{pmatrix},\\
    \widetilde{C}(z,x)&=\begin{pmatrix}
        -i\,\theta_r(-r\,x\,z) & \theta_r(-r/x\,z)\\
        +i\,\theta_r(+r\,x\,z) & \theta_r(+r/x\,z)\\
    \end{pmatrix}
    \begin{pmatrix}
        \frac{(1/x,-1/x:r)_\infty}{(-1,1/x^2;r)_\infty} & 0\\
        0 & \frac{(x,-x:r)_\infty}{(-1,x^2;r)_\infty}
    \end{pmatrix}.    
\end{align*}

The monodromy coordinates can now by computed directly. To this end, we note that $(x_1,x_2,x_3)=(r^{-1},r^{-2},r^{-3})$, so that
\begin{equation*}
    \rho_k=\rho_k(x)=\pi\left[C(r^{-k},t)\right]=(-1)^k \frac{\theta_r(-x)}{\theta_r(+x)},\qquad x=t^{\frac{1}{3}},
\end{equation*}
for $k=0,1,2$. In particular, we have
\begin{equation*}
    \rho_k(r\,x)=-\rho_k(x)=\rho_k(1/x),\quad \rho_k(-1/x)=-1/\rho_k(x),
\end{equation*}
which confirms that the coordinates satisfy the $q$-difference equation \eqref{eq:rho_qdif} as well as symmetries \eqref{eq:rho_sym} and \eqref{eq:rho_sym_plus}. Furthermore, we note that the monodromy coordinates have three branches in the complex $t$-plane, each corresponding to a particular branch of the solution $f$.

\begin{remark}Note that in light of Lemma \ref{lem:forbidden}, the only values of $x$ for which the coordinates lie on the curve \eqref{eq:forbidden_cubic}, are given by 
\begin{equation*}
    x=(-\tfrac{1}{2}\pm \tfrac{1}{2}\sqrt{3})r^n\qquad (n\in\mathbb{Z}),
\end{equation*}
which correspond to values of $t$ lying in $q^\mathbb{Z}$ and thus violate the non-resonance conditions \eqref{eq:conditionnonresonant}.
\end{remark}

\subsection{Solvable monodromy of the $q$-Okamoto rational solutions}
In this section, we consider how to generate the monodromy coordinates of the whole family of rational solutions in Theorem \ref{thm:oka_rational}. We do so by applying translation elements $T_{1,2,3}$ in the affine Weyl symmetry group $(A_2+A_1)^{(1)}$, see \cite{kajiwaranoumiyamada2001}, which act on the parameters as
\begin{align*}
    T_1:& \quad (a_0,a_1,a_2)\mapsto (q\, a_0, a_1/q,a_2),\\
    T_2:& \quad  (a_0,a_1,a_2)\mapsto (a_0, q\,a_1, a_2/q),\\
    T_3:& \quad  (a_0,a_1,a_2)\mapsto (a_0/q,  a_1,q\,a_2).
\end{align*}
It was shown in \cite{joshinobu2016} that these translations act as Schlesinger transformations on the spectral equation \eqref{eq:laxspectral}.

By methods similar to the derivation of equation \eqref{eq:rho_qdif}, it can be shown that these translations act on the monodromy coordinates as follows
   \begin{align*}
    T_1:& \quad (\rho_1,\rho_2,\rho_3)\mapsto (-\rho_1,-\rho_2,+\rho_3),\\
    T_2:& \quad (\rho_1,\rho_2,\rho_3)\mapsto (-\rho_1,+\rho_2,-\rho_3),\\
    T_3:& \quad (\rho_1,\rho_2,\rho_3)\mapsto (+\rho_1,-\rho_2,-\rho_3).
\end{align*} 

The family of rational solutions in Theorem \ref{thm:oka_rational} are indexed by $(m,n)\in\mathbb{Z}^2$. The translations act on the family of rational solutions through the following shifts of indices,
\begin{equation*}
    T_1:(m,n)\mapsto (m+1,n),\quad T_2:(m,n)\mapsto (m,n+1),\quad T_3:(m,n)\mapsto (m-1,n-1).
\end{equation*}
It follows that, for general $m,n\in\mathbb{Z}$, the monodromy coordinates corresponding to the rational solution in Theorem \ref{thm:oka_rational}, with indices $(m,n)$, are given by
\begin{align}
    &\rho_1(x)=(-1)^{1+m+n}s(x), \nonumber\\
    &\rho_2(x)=(-1)^{m}s(x), \qquad \hspace{1cm} s(x):=\frac{\theta_r(-x)}{\theta_r(+x)}.\label{eq:rhoexplicit}\\
    &\rho_3(x)=(-1)^{1+n}s(x). \nonumber
\end{align}

We proceed to check that these formulas are consistent with equation \eqref{eq:p_explicit} in Theorem \ref{thm:solvable}. Recalling equations \eqref{eq:initiali}, which provide the rational solutions at $x=-i$, we find the initial conditions at $t=i$:
\begin{equation*}
   (v_0,v_1,v_2)=(f_0(i),f_1(i),f_2(i))=((-1)^{1+m},(-1)^{1+m+n},(-1)^{1+n}).
\end{equation*}
Similarly, evaluating the expressions for the $\rho$-coordinates in equations \eqref{eq:rhoexplicit} at $x=-i$, leads to
\begin{equation*}
    (\rho_1(-i),\rho_2(-i),\rho_3(-i))=((-1)^{m+n}i,(-1)^{m+1}i,(-1)^n i).
\end{equation*}
These two expressions are consistent with equation \eqref{eq:p_explicit}.

\begin{figure}\captionsetup[subfigure]{labelformat=empty}
	\centering
\begin{subfigure}[b]{0.46\textwidth}
\centering
 \includegraphics[width=\textwidth]{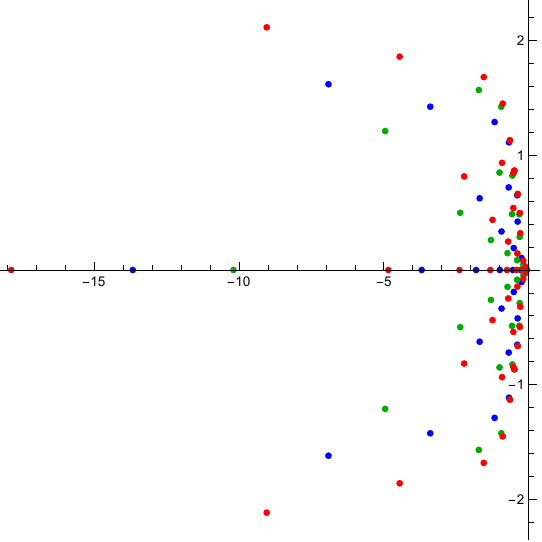}
 \caption{$k=3$}
 \end{subfigure}\hfill
 \begin{subfigure}[b]{0.46\textwidth} 
 \centering
 \includegraphics[width=\textwidth]{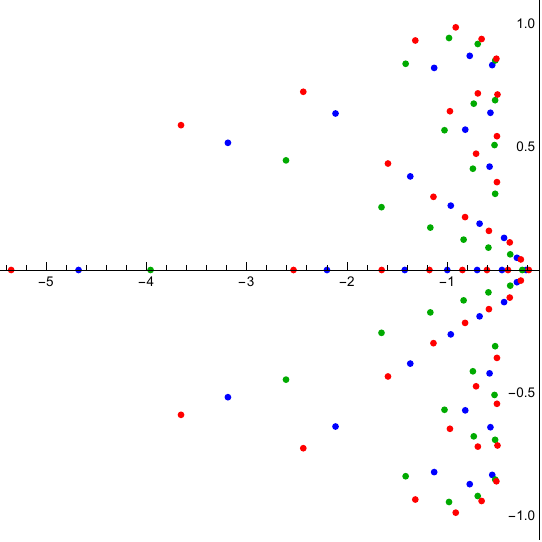} 
 \caption{$k=4$}
 \end{subfigure}
  \begin{subfigure}[b]{0.46\textwidth}
  \centering
 \includegraphics[width=\textwidth]{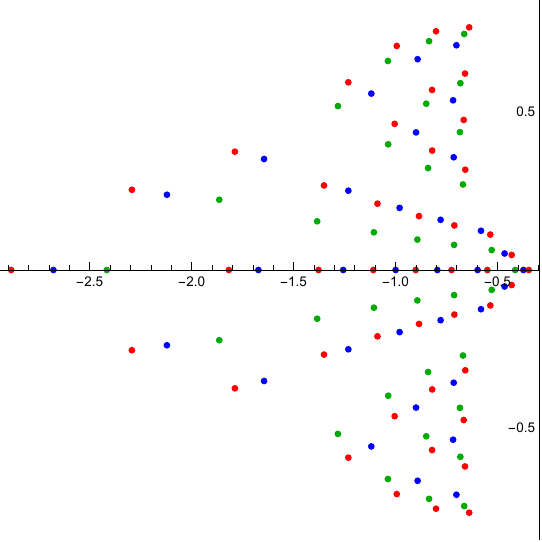}
 \caption{$k=5$}
 \end{subfigure}
 \hfill \begin{subfigure}[b]{0.46\textwidth}
 \centering
 \includegraphics[width=\textwidth]{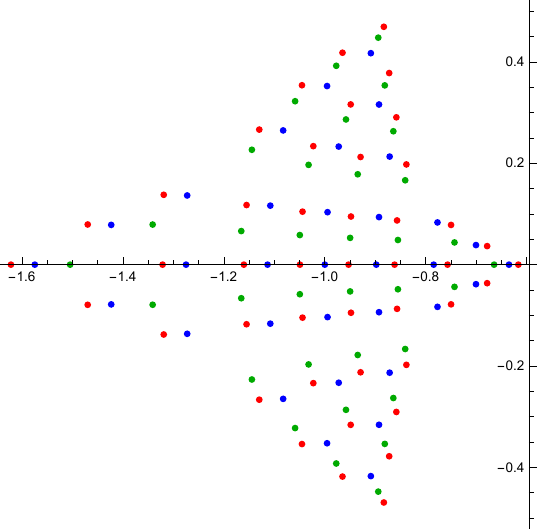}
  \caption{$k=7$}
 \end{subfigure}
  \begin{subfigure}[b]{0.46\textwidth}
  \centering
 \includegraphics[width=\textwidth]{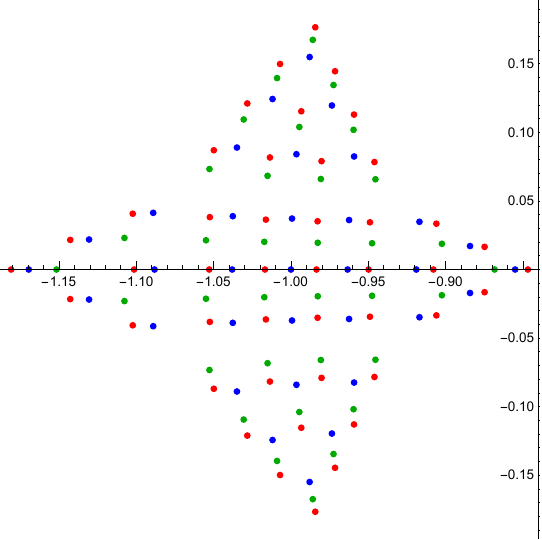}
  \caption{$k=10$}
 \end{subfigure} \hfill
 \begin{subfigure}[b]{0.46\textwidth}
 \centering
 \includegraphics[width=\textwidth]{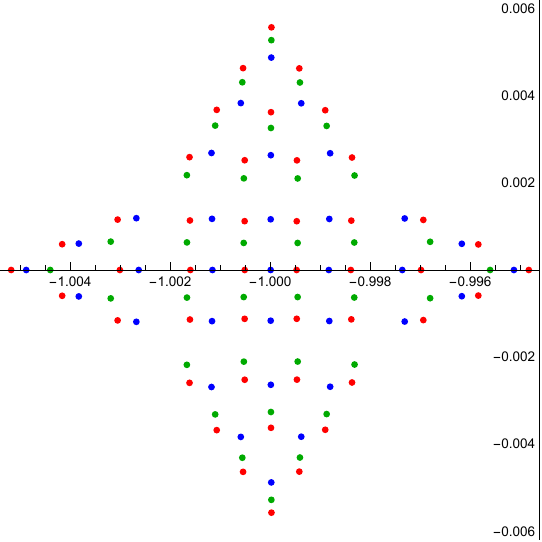}
  \caption{$k=20$}
 \end{subfigure}
\caption{In these plots the roots of the polynomials occurring in the definition of the $q$-Okamoto rational solution in Theorem \ref{thm:oka_rational}, with $(m,n)=(4,7)$, are displayed, where the value of $q=r^3$ varies between the plots by $r=1-(1/2)^k$, with $k=3,4,5,7,10,20$.
 In each figure, the blue, green and red dots represent zeros of $Q_{m,n}(x)$, $Q_{m+1,n}(x)$ and $Q_{m+1,n+1}(x)$ respectively. 
} \label{fig:pole_distributions}
\end{figure}

We conclude the section with some graphical representations of the pole distributions of a $q$-Okamoto rational solution in Figure \ref{fig:pole_distributions}. 

%% file: conclusion.tex
\section{Conclusion}\label{s:conc}
We have shown that two symmetries $\mathcal{T}_{\pm}$ of $q\Pfour$ can be lifted to the corresponding Lax pair and monodromy manifold. We have derived four symmetric solutions of $q\Pfour$ on the discrete time domain $q^\mathbb{Z}i$, which are invariant under $\mathcal{T}_{-}$.
We have further shown that they lead to solvable monodromy problems at the reflection point $t=i$, which provided an explicit correspondence between the four symmetric solutions and the four points on the monodromy manifold invariant under $\mathcal{T}_{-}$ in Theorem \ref{thm:solvable}.

We also studied the family of $q$-Okamoto rational solutions and showed that they are invariant under both $\mathcal{T}_{+}$ and $\mathcal{T}_{-}$. We further showed that their simplest member leads to an explicitly solvable monodromy problem in its entire $t$-domain. We used this to determine the values of the monodromy coordinates on the monodromy manifold for all the $q$-Okamoto rational solutions. The computation of the monodromy for the $q$-Okamoto rational solutions in Section \ref{s:mero} could serve as a starting point for deducing similar results for other $q$-equations. 

The pole distributions of the classical Okamoto rational solutions to $\Pfour$ have been analysed via Riemann-Hilbert methods \cite{buckmiller} and the Nevanlinna theory of branched coverings of the Riemann sphere \cite{masoeroroffelsen}. The extension of such studies to the $q$-difference Painlev\'e equations is an open problem.

The results of this paper yield Riemann-Hilbert representations for both the symmetric solutions on discrete time domains and the $q$-Okamoto rational solutions, through the theory set up in our previous paper \cite{joshiroffelsenrhp}. These can in turn form the basis of the rigorous asymptotic analysis of these solutions, as $t$ grows small or large or some of the parameters tend to infinity.

%% file: app_notation.tex
\section{Notation}\label{app:not}
Define the Pauli matrices
\begin{equation*}
\sigma_1=\begin{pmatrix}
0 & 1\\
1 & 0
\end{pmatrix},\quad
\sigma_2=\begin{pmatrix}
0 & -i\\
i & 0
\end{pmatrix},\quad
\sigma_3=\begin{pmatrix}
1 & 0\\
0 & -1
\end{pmatrix}.
\end{equation*}
We define the $\mathit{q}$-Pochhammer symbol by means of the infinite product
\begin{equation*}
(z;q)_\infty=\prod_{k=0}^{\infty}{(1-q^kz)}\qquad (z\in\mathbb{C}),
\end{equation*}
which converges locally uniformly in $z$ on $\mathbb{C}$. In particular $(z;q)_\infty$ is an entire function, satisfying
\begin{equation*}
(qz;q)_\infty=\frac{1}{1-z}(z;q)_\infty,
\end{equation*}
with $(0;q)_\infty=1$ and simple zeros on the semi $q$-spiral $q^{-\mathbb{N}}$. The $\mathit{q}$-theta function is defined as
\begin{equation}\label{eq:thetasym}
\theta_q(z)=(z;q)_\infty(q/z;q)_\infty\qquad (z\in \mathbb{C}^*),
\end{equation}
which is analytic on $\mathbb{C}^*$, with essential singularities at $z=0$ and $z=\infty$ and simple zeros on the $q$-spiral $q^\mathbb{Z}$. It satisfies
\begin{equation*}
\theta_q(qz)=-\frac{1}{z}\theta_q(z)=\theta_q(1/z).
\end{equation*}

For $n\in\mathbb{N}^*$ we denote
\begin{align*}
\theta_q(z_1,\ldots,z_n)&=\theta_q(z_1)\cdot \ldots\cdot \theta_q(z_n),\\
(z_1,\ldots,z_n;q)_\infty&=(z_1;q)_\infty\cdot\ldots\cdot (z_n;q)_\infty.
\end{align*}

For conciseness, we will use bars to denote iteration in $t$. That is, for $f=f(t)$, we denote $f(q\,t)=\overline f$, and $f(t/q)=\underline f$.

%% file: app_technical_lem.tex
\section{Proof of a technical lemma}\label{app:technical_lemma}

\begin{proof}[Proof of Lemma \ref{lem:forbidden}]
Let $C(z,t)$ be the connection matrix corresponding to the solution $f$. Let $t_*\in q^\mathbb{Z}t_0$ be such that $f(t_*)$ is regular. Then the Lax matrix $A(z,t_*)$ is well-defined at this point and consequently, we have a corresponding connection matrix $C(z,t_*)$ defined via equation \eqref{eq:assodefiiv}. Furthermore, using the time-evolution of the connection matrix in equation \eqref{eq:connection_evolution}, we can thus infer that $C(z,t_0)$ is also well-defined.

Now suppose, on the contrary, that the corresponding monodromy coordinates,
\begin{equation*}
    p_k=\pi(C(x_k,t_0)),
\end{equation*}
lie on the curve defined by the cubic equations \eqref{eq:forbidden_cubic}. We are going to obtain a contradiction by showing that $C(z,t_0)$ does not satisfy property \ref{item:c3}.
To this end, we will first obtain a general parametrisation of this curve.

Consider the following matrix function,
\begin{equation}\label{eq:psuedo_connect}
    \mathcal{C}(z)=\begin{pmatrix}
    C_1(z) & C_2(z)\\
    -C_1(-z) & C_2(-z)
    \end{pmatrix},
\end{equation}
where
\begin{align*}
    C_1(z)&=\theta_q(+z/u,-z/u,z/w), & u^2 w=&\frac{1}{q a_0^2 a_2} t^{-1},\\
    C_2(z)&=z\theta_q(+z/v,-z/v,z/w), & qv^2 w=&\frac{1}{q a_0^2 a_2} t^{+1},\\
\end{align*}
for any choice of $t,w\in\mathbb{C}^*$. This matrix satisfies properties \ref{item:c1}, \ref{item:c2}, \ref{item:c4}, as well as a degenerate version of \ref{item:c3}, namely
\begin{equation*}
    |\mathcal{C}(z)|\equiv 0.
\end{equation*}

The monodromy coordinates, $P_k=\pi(\mathcal{C}(x_k))$, $k=1,2,3$, of this pseudo-connection matrix, read
\begin{equation}\label{eq:parametrisation_forbidden}
    (P_1,P_2,P_3)=\bigg(-\frac{\theta_q(+x_1/w)}{\theta_q(-x_1/w)},-\frac{\theta_q(+x_2/w)}{\theta_q(-x_2/w)},-\frac{\theta_q(+x_3/w)}{\theta_q(-x_3/w)}\bigg).
\end{equation}
These monodromy coordinates solve the cubic \eqref{eq:cubic} and their expressions are completely independent of $t$. In other words, they lie on the intersection of cubics \eqref{eq:cubic}, as $t$ varies in $\mathbb{C}^*$. In particular, these monodromy coordinates must lie on the curve defined by \eqref{eq:forbidden_cubic}.

We will show that \eqref{eq:parametrisation_forbidden} completely parametrises the curve defined by \eqref{eq:forbidden_cubic}, as $w$ varies in $\mathbb{C}^*$. Since we have not assumed anything on $(p_1,p_2,p_3)$, this is equivalent to proving that there exists a $w$ such that
\begin{equation}\label{eq:pequality}
    (P_1,P_2,P_3)=(p_1,p_2,p_3).
\end{equation}

Now, the equation
\begin{equation*}
    p_1=-\frac{\theta_q(+x_1/w)}{\theta_q(-x_1/w)},
\end{equation*}
has two, counting multiplicity, solutions $w_{1,2}$, on the elliptic curve $\mathbb{C}^*/q^2$, related by
$w_2\equiv q x_1^2/w_1$ modulo multiplication by $q^2$.

For either choice, $w=w_1$ or $w=w_2$, we have $p_1=P_1$ and the pairs $(P_2,P_3)$ and $(p_2,p_3)$ satisfy the same two equations \eqref{eq:forbidden_cubic}, which are quadratic in the remaining variables. In fact, upon fixing the value of $p_1$, \eqref{eq:forbidden_cubic} has two solutions (counting multiplicity), and these two solutions coincide if and only if $w_1$ and $w_2$ coincide on the elliptic curve $\mathbb{C}^*/q^2$. It follows that \eqref{eq:pequality} holds for $w=w_1$ or $w=w_2$.

We now fix $w$ such that \eqref{eq:pequality} holds, set $t=t_0$ in \eqref{eq:psuedo_connect}, and consider the quotient
\begin{equation*}
    D(z)=C(z,t_0)^{-1}\mathcal{C}(z).
\end{equation*}
Since $C(z,t_0)$ and $\mathcal{C}(z)$ have the same monodromy-coordinate values, $D(z)$ is analytic at $z=\pm x_k$, $k=1,2,3$ and thus forms an analytic matrix function on $\mathbb{C}^*$. Then, by property $\ref{item:c2}$,
\begin{equation*}
    D(qz)=t_0^{\sigma_3}D(z)t_0^{-\sigma_3}.
\end{equation*}
Since $t_0^2\notin q^\mathbb{Z}$, the only analytic matrix functions satisfying this $q$-difference equation are constant diagonal matrices, and therefore $D$ is simply a constant diagonal matrix. But then
\begin{equation*}
    C(z,t_0)D=\mathcal{C}(z),
\end{equation*}
and neither diagonal entry of $D$ can equal zero, as this contradicts equation \eqref{eq:psuedo_connect}, so $|D|\neq 0$. Hence
\begin{equation*}
    |C(z,t_0)|=|\mathcal{C}(z)|/|D|\equiv 0,
\end{equation*}
which contradicts property \ref{item:c3}. The lemma follows.
\end{proof}